\documentclass[journal=jctcce,manuscript=article]{achemso}

    \usepackage[pdftex,bookmarks=false]{hyperref}
    \usepackage{amssymb, amsmath}
    \usepackage{graphicx}
    \usepackage{bm}
    \usepackage{makecell}
    \usepackage{siunitx}
    
    \graphicspath{Figures}

\author{N.~Unglert}
\affiliation{Institute of Materials Chemistry, TU Wien, 1060 Vienna, Austria} 
\author{L. B. Pártay}
\affiliation{Department of Chemistry, University of Warwick, Coventry CV4 7AL, UK}
\author{G. K. H. Madsen}
\email{georg.madsen@tuwien.ac.at}
\affiliation{Institute of Materials Chemistry, TU Wien, 1060 Vienna, Austria}

\title{Replica exchange nested sampling}

\begin{document}

    \begin{abstract}
        Nested sampling (NS) has emerged as a powerful tool for exploring thermodynamic properties in materials science. However, its efficiency is often hindered by the limitations of Markov chain Monte Carlo (MCMC) sampling. In strongly multimodal landscapes, MCMC struggles to traverse energy barriers, leading to biased sampling and reduced accuracy. To address this issue, we introduce replica-exchange nested sampling (RENS), a novel enhancement that integrates replica-exchange moves into the NS framework. Inspired by Hamiltonian replica exchange methods, RENS connects independent NS simulations performed under different external conditions, facilitating ergodic sampling and significantly improving computational efficiency.
        We demonstrate the effectiveness of RENS using four test systems of increasing complexity: a one-dimensional toy system, periodic Lennard-Jones, the two-scale core-softened Jagla model and a machine learned interatomic potential for silicon. Our results show that RENS not only accelerates convergence but also allows the effective handling of challenging cases where independent NS fails, thereby expanding the applicability of NS to more realistic material models.
    \end{abstract}

    \maketitle

\section{Introduction}
Nested sampling (NS) has emerged as a powerful method for simulating thermodynamic properties in materials science \cite{partay_nested_2021}. 
Nested sampling provides an unbiased exploration of the configuration space, functioning as a top-down exploration method. It is an iterative technique that partitions the configuration space into a nested sequence of phase space volumes, each confined by iso-likelihood surfaces. In each iteration, a layer of this sequence is peeled away, producing a corresponding sample and assigning it a specific phase-space volume. The core challenge in NS is the generation of new samples within these confined volumes. In the language of Bayesian statistics, this process is referred to as likelihood-constrained-prior sampling and is typically performed through a Markov chain Monte Carlo (MCMC) random walk \cite{skilling_bayesian_2012, griffiths_nested_2019}. In this framework, new samples are generated by cloning an existing point, known as a walker, and performing a decorrelating random walk. 
To ensure ergodicity, the MCMC procedure must be capable of exploring all relevant regions of the configuration space, which can be a challenge. 

Nested sampling procedures mitigate some of the inherent MCMC limitations by maintaining a collection of walkers, each representing a sample from the likelihood-constrained prior at a given iteration. A larger number of walkers serves two key purposes. First, maintaining a diverse set of walkers increases the probability of capturing all significant modes of the posterior distribution, thereby compensating for the failure of individual Markov chains to fully explore the configuration space. Second, it enables finer-grained sampling of the parameter space, reducing the shrinkage of the sampled region at each iteration and improving the precision of the numerical integration of the evidence integral.

While early applications of NS relied on simple Hamiltonians or semi-empirical potentials, recent advances in machine-learned force fields (MLFFs) have extended the application of NS to more realistic systems \cite{marchant_exploring_2023, kloppenburg_general-purpose_2023, rosenbrock_machine-learned_2021, unglert_neural-network_2023}. As the complexity of the studied systems increases, the potential energy surfaces that need to be explored also become increasingly challenging for the underlying MCMC algorithm.
Achieving ergodicity for high-dimensional multimodal potential energy surfaces requires a large number of walkers, often in the order of a few 1000s \cite{unglert_neural-network_2023, marchant_exploring_2023, kloppenburg_general-purpose_2023}. The numerous local minima, each representing configurations with distinct properties, often require a prohibitive number of MCMC steps to traverse narrow regions of the parameter space. 
Furthermore, atomistic simulations often focus on observables under varying external conditions. Although the NS approach inherently accounts for temperature dependence, separate NS simulations are typically required for a set of external conditions such as pressure or chemical potential.

In this study, we present a replica exchange nested sampling (RENS) which links previously independent NS simulations, each conducted under different external conditions, together. 
The replica exchange (RE) method was originally proposed by Swendsen in 1986 \cite{swendsen_replica_1986} to address the problem of broken ergodicity in low-temperature spin glass simulations. This approach, also known as parallel tempering, involves simultaneously simulating a system at several different temperatures.
Each replica samples from a canonical distribution and periodically, after propagating the replicas independently for a fixed number of steps, an exchange attempt is made between pairs of replicas with adjacent temperatures, based on an appropriate acceptance criterion.
Several strategies have been proposed for the optimal selection of replicas and the self-adaptive adjustment of replica temperatures, enhancing convergence \cite{mandelshtam_structural_2006, miasojedow_adaptive_2013,rozada_effects_2019}. 
The concept of RE has been extended beyond temperature-based replicas to include replicas with different interaction potentials, an approach known as Hamiltonian RE \cite{sugita_multidimensional_2000}. Hamiltonian RE has become widely adopted in the molecular simulation community, particularly for studying biomolecular systems \cite{fukunishi_hamiltonian_2002}.
Recent developments have further expanded the versatility of the RE method, combining it with techniques such as umbrella sampling \cite{park_transmembrane_2012, oshima_replica-exchange_2019}, extended ensembles \cite{hsu_replica_2024} and transition path sampling \cite{falkner_enhanced_2024}, among others.

Like the earlier superposition enhanced NS \cite{martiniani_superposition_2014}, our proposed RENS method can be viewed as a specific implementation of Hamiltonian RE for NS. Both approaches share the key feature of generating samples from 
multiple likelihood-constrained prior distributions, which constitute the fundamental quantities to be sampled within the NS framework.

In this work, we use four different model systems of increasing complexity to demonstrate how the proposed RENS method not only significantly enhances sampling efficiency of MCMC, but also allows the effective handling of scenarios where independent NS fails.  We start by explaining and illustrating the method using a simple one-dimensional toy model and the periodic Lennard-Jones system. We then apply our approach to two challenging cases where traditional independent NS has previously struggled to predict correct thermodynamic behavior, namely the Jagla model \cite{hemmer_fluids_1970,jagla_liquid-liquid_2001,bartok_insight_2021} and the silicon phase diagram \cite{unglert_neural-network_2023}.

\section{Methods}

\subsection{Nested sampling}

Nested sampling is widely employed as a tool for Bayesian inference \cite{skilling_nested_2006,ashton_nested_2022}, as it provides an approximation of the evidence, which appears in the well-known Bayes' theorem
\begin{align}
    p(\theta) = \frac{L(\theta) \; \pi(\theta)}{Z},
    \label{eq:bayes_general}
\end{align}
where \( p \) represents the posterior, \( L \) the likelihood, \( \pi \) the prior, and \( Z \) the evidence for a parameter space $\theta$. 

In the context of atomistic systems at constant pressure, this framework is analogous to
\begin{align}
    p(\bm{r}) = \frac{\exp[-\beta H(\bm{r})] \; f(\bm{r})}{Q_{NPT}},
    \label{eq:bayes_npt}
\end{align}
where the posterior corresponds to the distribution of the isothermal-isobaric ensemble, normalized by the isothermal-isobaric partition function \( Q_{NPT} \), where $N$, $P$ and $T$ are the number of particles, pressure and temperature, respectively. The likelihood is expressed as the Boltzmann factor, which depends on the inverse temperature $\beta=(k_\mathrm{B} T)^{-1}$ and the microscopic enthalpy \( H \) as a function of the configuration \( \bm{r} \). Here, $\bm{r}$ contains the complete structural information of the system, which, in the case of periodic boundary conditions, can be decomposed into the Cartesian coordinates of a repeating unit and the simulation cell.
In this framework, the prior distribution, $f$, is uniform, reflecting the principle of equal {\it a priori} probability. For simplicity, we consider only the configurational part of the distribution, as it can be straightforwardly decoupled from the momentum-dependent part in accordance with classical statistical mechanics.

The core principle of the NS algorithm lies in the generation of samples from progressively thinner nested shells within the parameter space. The likelihood of each consecutive sample determines a volume within the parameter space, which is enclosed by the volume defined by the previous sample.
This iterative reduction of the accessible phase space volume is facilitated by maintaining a pool of $K$ walkers. At each iteration, $i$, the walker with the highest enthalpy, $H_i^\mathrm{lim}$, determines the iso-likelihood hypersurface that defines the prior constraint for that step. This walker is then removed from the pool and stored for post-processing, enabling the evaluation of the partition function later on. Subsequently, a new walker is generated by sampling from the prior distribution, now restricted by the corresponding likelihood threshold.
The choice of the prior distribution is a fundamental degree of freedom in Bayesian inference, representing the extent of prior knowledge incorporated into the estimation of the posterior distribution.
In the context of atomistic systems, the most straightforward choice is an uninformed uniform prior. At a given iteration $i$, the probability density of the walkers' distribution can then be expressed as the likelihood-constrained prior distribution:
\begin{align}
    f^{\mathrm{uniform}}_i(\bm{r}) &= 
        \begin{cases}
        \Delta_i^{-1} & \text{if } \; H(\bm{r}) < H_i^{\mathrm{lim}} \\
        0 & \text{otherwise}
        \end{cases}.
    \label{eq:uniform_prior}
\end{align}
Again, $H^{\mathrm{lim}}_i$ is the likelihood threshold for the current iteration, and $\Delta_i$ is the normalization constant related to the integrated density of states.


To improve sampling efficiency, the volume distribution from the isothermal-isobaric ensemble can be incorporated into a volume informed prior, which is straightforward in an atomistic Monte Carlo simulation. The volume-aware constrained prior density can then be defined as:
\begin{align}
    f^{\mathrm{vol}}_i(\bm{r}) &= 
        \begin{cases}
        \frac{V(\bm{r})^N}{\Delta_i'} & \text{if } \; H(\bm{r}) < H_i^{\mathrm{lim}} \\
        0 & \text{otherwise}
        \end{cases}.
    \label{eq:vol_prior}
\end{align}
The NS formalism is particularly elegant due to the nature of the samples it generates. 
The algorithm delivers an estimate for the parameter space volume corresponding to each sample, which serve as weights $w_i$ in a numerical approximation of the partition function from Eq.~\eqref{eq:bayes_npt}.
These samples can be interpreted as weighted, allowing for the approximation of the partition function at any temperature through
\begin{align} 
    Q_{NPT}(\beta) &= \int \mathrm{d} \bm{r} e^{-\beta H(\bm{r})} \approx \sum_{i} w_i e^{-\beta H(\bm{r})}.
    \label{eq:ns_partition_func}
\end{align}
where the weights $w_i$ arise directly from the NS algorithm.
By construction, the $w_i$ are of statistical nature \cite{skilling_nested_2006} and can be estimated as $w_i = \overline{X_{i-1}} - \overline{X_{i}}$. Here, $\overline{X_{i}} = X_0 [K/(K + 1)]^i$ is the average enclosed prior mass at iteration $i$, which in the atomistic sense corresponds to the configuration space volume enclosed by the $i$-th sample.

The computational bottleneck of NS procedures is the acquisition of samples from the likelihood-constrained prior distributions. 
To date, there exist essentially two methods \cite{ashton_nested_2022} for this task: region samplers and MCMC samplers. 
Region samplers \cite{feroz_multinest_2009} construct a geometric shape enclosing the likelihood contour, subsequently performing rejection sampling based on independent and identically distributed (\textit{iid}) samples drawn from within the shape. If the shape is always guaranteed to contain the entire volume enclosed by the likelihood threshold, this procedure yields exact samples of the target distribution. Unfortunately, region samplers are limited to low-dimensional problems in practice due to the curse of dimensionality.
In contrast, MCMC samplers can handle high-dimensional parameter spaces by cloning one of the already existing walkers and performing a random walk until the clone is decorrelated and can be regarded as a new \textit{iid} sample from the constrained prior. In the limit of infinite walk length, this is also guaranteed to yield an exact sample of the target distribution. However, in practice the required walk lengths are in many scenarios prohibitive and lead to biased samples due to the Markov chain being trapped in certain modes of the distribution.

\subsection{Replica-exchange nested sampling}

By analogy with the derivation of the acceptance criterion for the Hamiltonian RE algorithm \cite{sugita_multidimensional_2000}, we derive an acceptance criterion for the RE procedure used in RENS. We perform the derivation for NS replicas simulating isobaric-isothermal ensembles at different pressures. However, we emphasize that the developed formalism is general and can be applied to any set of $M$ distributions with probability densities \{$p^{[1]}(\theta)$,~\dots,~ $p^{[M]}(\theta)$\} that operate within the same parameter space $\theta$.
In the case of constant volume parallel tempering, the target distributions are the canonical distributions at different temperatures 
$p^{[m]}(\bm{r}) \propto \exp[-\beta^{[m]} U(\bm{r})]$. 
For RENS the target distributions are the likelihood constrained prior distributions for different simulations $p^{[m]}(\bm{r}) = f_i^{[m]}(\bm{r})$ at the same iteration $i$. Here,
\begin{align}
    f_i^{[m]}(\bm{r}) &= 
        \begin{cases}
        \frac{A(\bm{r})}{\Delta_i^{[m]}} & \text{if } \; H^{[m]}(\bm{r}) < H_i^{[m],\mathrm{lim}} \\
        0 & \text{otherwise}
        \end{cases},
    \label{eq:prior_general}
\end{align}
which is a generalization of Eq.~\eqref{eq:uniform_prior} and \eqref{eq:vol_prior}, now introducing also the possibility for different enthalpy functions $H^{[m]}$ as well as an additional arbitrary dependence on $\bm{r}$ through $A(\bm{r})$.
At a given iteration $i$ of a RENS simulation consisting of $M$ replicas, we can define a generalized ensemble with probability density
\begin{align}
    W_i(R_i) &= \prod_{m}^{M} f_i^{[m]}(\bm{r}_i^{[m]}).
    \label{eq:generalized_ensemble}
\end{align}
Here, $R_i = \{\bm{r}_i^{[1]},...,\bm{r}_i^{[M]}\}$ corresponds to a single "microstate" in this generalized ensemble, where the replicas are labeled by $m=\{1, \dots, M\}$. $\bm{r}_i^{[m]}$ is thus a configuration from replica $m$ at iteration $i$ and $f_i^{[m]}(\bm{r})$ is the corresponding likelihood-constrained prior density. Note, that $f_i^{[m]}$ changes with each iteration $i$. 
In order to satisfy detailed balance, the acceptance probability has to fulfill
\begin{align}
    \mathcal{P}_i^\mathrm{acc}(R_i'\,|\,R_i) &= \min \bigg[ 1,  \frac{W_i(R_i')}{W_i(R_i)} \bigg],
    \label{eq:acc_general}
\end{align}
where $R_i$ and $R_i'$ are two arbitrary microstates in the generalized ensemble. If we choose $R_i'$ and $R_i$ to be different only by a single swap move between two neighboring replicas $s$ and $t$, such that 
$R_i = \{\bm{r}_i^{(1)}, \dots, \bm{r}_i^{[s]} , \bm{r}_i^{[t]}, \dots ,\bm{r}_i^{(M)}\}$ 
and 
$R_i'= R^{s \leftrightarrow t}_i = \{\bm{r}_i^{(1)}, \dots, \bm{r}_i^{[t]} , \bm{r}_i^{[s]}, \dots ,\bm{r}_i^{(M)}\}$,
we can derive an explicit expression for the acceptance criterion of this swap move.
Using the likelihood constrained prior distribution $f_i^{[m]}(\bm{r})$ as well as Eq.~\eqref{eq:generalized_ensemble}, we can rewrite the $W_i$ ratio from Eq.~\eqref{eq:acc_general}.
As a consequence of considering only a single swap ($s \leftrightarrow t$), all other terms cancel and we get

\begin{align}
\begin{split}
    \frac{W_i(R^{s \leftrightarrow t}_i)}{W_i(R_i)} 
    &= \frac{
            f_i^{[s]}(\bm{r}_i^{[t]}) \; f_i^{[t]}(\bm{r}_i^{[s]})
        }{
            f_i^{[s]}(\bm{r}_i^{[s]}) \; f_i^{[t]}(\bm{r}_i^{[t]})
        } \\
    &= \begin{cases}
        1   
            & \text{if} \quad 
                H^{[s]}(\bm{r}_i^{[t]}) < H_i^{[s], \mathrm{lim}} \quad \text{and } \\
            & \phantom{\text{if} \quad } 
                 H^{[t]}(\bm{r}_i^{[s]}) < H_i^{[t], \mathrm{lim}} \\
        0 & 
            \text{otherwise}.
    \end{cases}
\end{split}
\label{eq:pacc_enthalpy}
\end{align}

The microscopic enthalpy is computed according to 

\begin{align}
    H^{[m]}(\bm{r}) = U(\bm{r}) + P^{[m]} V(\bm{r}),
    \label{eq:H_m}
\end{align}
with the volume function $V(\bm{r})$ and pressure $P^{[m]}$ of replica $m$.
$H_i^{[m], \mathrm{lim}}$ is the enthalpy limit in iteration $i$ for replica $m$. 
Note, that the choice of the prior distribution (c.f. Eq.~\eqref{eq:prior_general}) does not alter Eq.~\eqref{eq:pacc_enthalpy}, since
\begin{equation*}
\begin{split}
    \frac{
        f_i^{[s]}(\bm{r}_i^{[t]}) \; f_i^{[t]}(\bm{r}_i^{[s]})
    }{
        f_i^{[s]}(\bm{r}_i^{[s]}) \; f_i^{[t]}(\bm{r}_i^{[t]})
    }
    = \frac{
        \frac{1}{\Delta_i^{[s]}} \; \frac{1}{\Delta_i^{[t]}}
    }{
        \frac{1}{\Delta_i^{[s]}} \; \frac{1}{\Delta_i^{[t]}}
    }
    = \frac{
        \frac{V_t^N}{{\Delta_i^{[s]}}'} \; \frac{V_s^N}{{\Delta_i^{[t]}}'}
    }{
        \frac{V_s^N}{{\Delta_i^{[s]}}'} \; \frac{V_t^N}{{\Delta_i^{[t]}}'}
    } 
    = 1 \\
    \text{for } \; H^{[s]}(\bm{r}_i^{[t]}) < H_i^{[s], \mathrm{lim}} 
        \quad \text{and} \quad H^{[t]}(\bm{r}_i^{[s]}) < H_i^{[t], \mathrm{lim}}
\end{split}
\end{equation*}
This leads to the expression for the acceptance rate
\begin{align}
    \mathcal{P}_i^\mathrm{acc}(R^{s \leftrightarrow t}_i\,|\,R_i) 
    = \begin{cases}
        1   
            & \text{if} \quad 
                H^{[s]}(\bm{r}_i^{[t]}) < H_i^{[s], \mathrm{lim}}  \\
            & \text{ and }
                 H^{[t]}(\bm{r}_i^{[s]}) < H_i^{[t], \mathrm{lim}} \\
        0 & 
            \text{otherwise}.
    \end{cases}
\end{align}
Since the acceptance probability for such a single swap move is independent from all replicas except $s$ and $t$, we can express it as

\begin{align}
    \mathcal{P}_{i, s \leftrightarrow t}^\mathrm{acc}(\bm{r}_i^{[s]}, \bm{r}_i^{[t]}) &= 
        \begin{cases}
            1   
                & \text{if} \quad 
                    U(\bm{r}_i^{[t]}) + P^{[s]} V(\bm{r}_i^{[t]}) < H_i^{[s],\mathrm{lim}} \\   
                & \text{ and  } 
                    U(\bm{r}_i^{[s]}) + P^{[t]} V(\bm{r}_i^{[s]}) < H_i^{[t],\mathrm{lim}} \\
            0 & 
                \text{otherwise},
        \end{cases}
\label{eq:re_acc}
\end{align}
where we also inserted Eq.~\eqref{eq:H_m} to express the enthalpy.
Note that, since the enthalpies $H^{[s]}(\bm{r}_i^{[s]})$, $H^{[t]}(\bm{r}_i^{[t]})$ and thus also the potential energies $U(\bm{r}_i^{[s]})$, $U(\bm{r}_i^{[t]})$ for both configurations $\bm{r}_i^{[s]}$ and $\bm{r}_i^{[t]}$ are already computed in the course of the MCMC sampling, the RE steps require only a reevaluation of the $PV$ terms, which is in general extremely fast computationally.

\begin{figure} 
    \centering
    \includegraphics[width=1.\columnwidth]{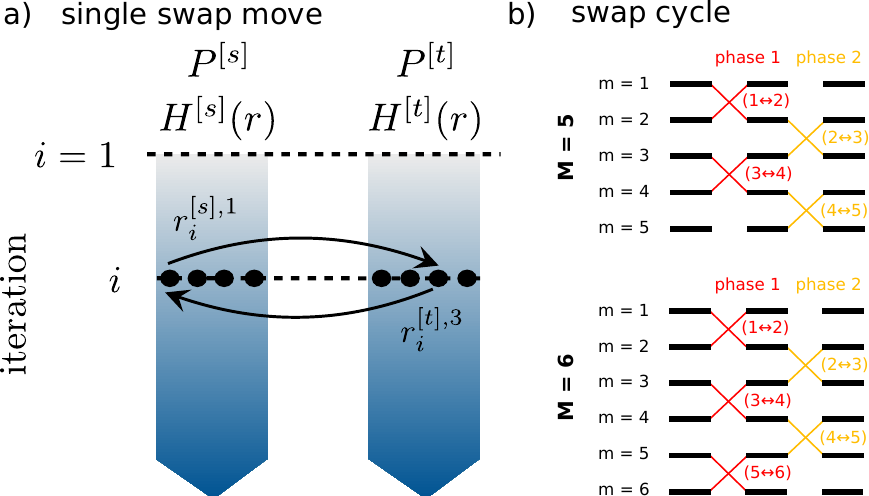}
    \caption{a) Schematic illustration of a single swap move between two replicas $s$ and $t$, run at pressures $P_s$ and $P_t$, respectively, at iteration $i$. Black points indicate four walker configurations, highlighting the randomly selected swap targets $r^{[s], k}_i$ and $r^{[t], k'}_i$. The downward blue arrow represents the direction of the sampling towards higher likelihood values.
    b) Schematic of the employed neighbor-based RE cycle, consisting of two swap phases. We show an example for a even and a odd number of replicas $M$, respectively.
    }
    \label{fig:swap_phases}
\end{figure}

In contrast to 
parallel tempering, each NS replica naturally maintains a set of samples 
$\{\bm{r}_i^{[m], k} \}$
from $f_i^{[m]}$, which are the walker configurations with $k=\{1, \dots, K\}$.
Fig.~\ref{fig:swap_phases}a shows the involved walker configurations schematically for a particular realization of a swap move between two replicas $s$ and $t$.
From Eq.~\eqref{eq:re_acc} it follows that two swap moves, $s \leftrightarrow t$ and $u \leftrightarrow v$, are statistically independent if $s,t,u$ and $v$ are unique. This can be expressed formally as
\begin{align}
    \mathcal{P}_{i, s \leftrightarrow t}^\mathrm{acc}
        \perp 
        \mathcal{P}_{i, u \leftrightarrow v}^\mathrm{acc}
        \quad \text{if} \quad
        \{u,v\}\cap \{s,t\} = \emptyset,
    \label{eq:independence}
\end{align}
where we dropped the arguments for the acceptance probability, since they are implicitly contained in the label.
This allows multiple swap attempts to be performed in parallel each time the RE mechanism is invoked. We restrict swap attempts to neighboring replicas (i.e. similar pressures) to ensure reasonable acceptance rates.

For practical purposes, we define two swap phases when simulating $M$ replicas, where for even $M$ the corresponding swap pairs are given by 
\begin{align*}
    \{ (m  \leftrightarrow m+1) \mid m=1, 3, \dots, M-2\} &\quad \text{phase 1}\\
    \{ (l  \leftrightarrow l+1) \mid l=2, 4, \dots, M-1\} &\quad \text{phase 2}
\end{align*}
and for odd $M$ by
\begin{align*}
    \{ (m  \leftrightarrow m+1) \mid m=1, 3, \dots, M-1\} &\quad \text{phase 1}\\
    \{ (l  \leftrightarrow l+1) \mid l=2, 4, \dots, M-2\} &\quad \text{phase 2}.
\end{align*}
Fig.~\ref{fig:swap_phases}b schematically illustrates the two swap phases for $M=5$ and $M=6$. 
At the beginning of each phase, two walkers are randomly selected from the walker populations of each swap pair's corresponding replicas, followed by an attempt to swap them.
This approach ensures that the statistical properties of the underlying distributions are maintained while maximizing parallelism.

In the case of RENS, multiple swap phases can also be executed sequentially. We define a RE call as a sequence of $n_\mathrm{cycles}$ RE cycles, where each cycle consists of two phases (phase 1 and phase 2).
This multi-cycle approach is particularly beneficial when the exchange mechanism introduces substantial overhead, such as in MPI-based communication. By performing multiple swap attempts sequentially, a sufficient mixing of distributions can be achieved while minimizing the frequency of RE calls.

\setlength{\tabcolsep}{5pt}
\begin{table*}[hbtp]
    \centering
    \begin{tabular}{c c c c c} 
        \hline\hline \\[-1.5ex]
            parameter           &   Toy     &   LJ      &   Jagla   &   Silicon \\ [0.5ex] 
            \hline 
            code    &
                                \texttt{cppnest} &
                                \texttt{cppnest} &
                                \texttt{cppnest} &
                                \texttt{JAXNEST} 
                                \\
            prior    &
                                uniform (Eq.~\eqref{eq:uniform_prior}) &
                                volume (Eq.~\eqref{eq:vol_prior}) &
                                volume &
                                volume 
                                \\
            walker init.    &  
                                random biased &  
                                cubic, grid & 
                                cubic, grid & 
                                triclinic, grid
                                \\
            $V_\mathrm{min}$    &  
                                \SI{0.25}{d \; \mathrm{atom}^{-1}} &  
                                \SI{0.5}{\sigma^3 \; \mathrm{atom}^{-1}} & 
                                \SI{0.5}{r_0^3 \; \mathrm{atom}^{-1}} & 
                                \SI{10.}{\angstrom^3 \; \mathrm{atom}^{-1}}
                                \\
            $V_\mathrm{max}$    &
                                \SI{5}{d \; \mathrm{atom}^{-1}} &  
                                \SI{100}{\sigma^3 \; \mathrm{atom}^{-1}} &
                                \SI{100}{r_0^3 \; \mathrm{atom}^{-1}} &
                                \SI{52.7}{\angstrom^3 \; \mathrm{atom}^{-1}}
                                \\
            $P_\mathrm{acc}$ window    &
                                (0.2, 0.5) &
                                (0.2, 0.5) &
                                (0.2, 0.5) &
                                (0.25, 0.75)
                                \\
            $N_\mathrm{adjust}$ &
                                100 &
                                100 &
                                100 &
                                400 
                                \\
            $f_\mathrm{adjust}$ &
                                1.5 &
                                1.5 &
                                1.5 &
                                1.5 
                                \\
            step types          &
                                \makecell{distance \\ lattice} &
                                \makecell{SP-MC\\ volume\\ stretch\\ shear} &
                                \makecell{SP-MC\\ volume\\ stretch\\ shear} &
                                \makecell{GMC\\ AP-MC\\ volume\\ stretch\\ shear} 
                                \\
            step ratio          &
                                1:1 &
                                1:10:1:1 &
                                8:4:2:2 &
                                1:8:16:8:8
                                \\
            $d_0$               &
                                - &
                                0.9 &
                                0.9 &
                                0.8
                                \\[1ex]
            $N_\mathrm{particles}$ &
                                2 &
                                64 &
                                64 &
                                16
                                \\[1ex]
         \hline\hline
        \end{tabular}
    \caption{Summary of important computational parameters employed throughout the simulations for the different investigated systems.}
    \label{tab:comp_params}
\end{table*}

\subsection{Computational details}

MCMC random walks often require hundreds of Monte Carlo steps to achieve convergence to the stationary distribution. Instead of performing a single, extended random walk of length $L$ on a cloned walker, a common parallelization strategy~\cite{baldock_constant-pressure_2017} selects a random subset of the remaining walkers and executes shorter random walks of length $L'$ concurrently across these configurations. This approach ensures that the expected total walk length, $\langle L \rangle$, for a walker before removal remains equal to $L$. We apply this parallelization strategy in all our simulations.

For constant-pressure NS we use atom moves and cell shape moves \cite{baldock_constant-pressure_2017}. The step sizes for each move type are dynamically adjusted by selecting a subset of the current walker population of size $N_\mathrm{adjust}$, generating trial moves for temporary clones of these configurations, and iteratively tuning the step sizes until the acceptance rate falls within a predefined range. Step size adjustments are performed at fixed intervals depending on $K$, with a minimum frequency of once every 100 iterations.

The applied cell moves include volume, stretch, and shear steps. Volume steps incorporate a rejection sampling procedure that accounts for the $V^N$ proportionality of the prior distribution, see Eq.~\eqref{eq:vol_prior}. The accessible configuration space is constrained by rejecting cell moves that yield volumes outside the interval $[V_\mathrm{min}, V_\mathrm{max}]$ or result in excessively skewed cells with a minimum aspect ratio \cite{baldock_determining_2016} smaller than $d_0$.

To decorrelate the positional degrees of freedom, we apply different types of atomic moves. For pair potentials, single-particle Monte Carlo (SP-MC) moves provide an efficient approach, where each SP-MC iteration attempts to displace every particle once with displacements sampled from a 3D Gaussian distribution. If forces are available, the Galilean MC algorithm \cite{skilling_bayesian_2012,skilling_galilean_2019, baldock_constant-pressure_2017} is well-suited for constrained prior volumes and we execute it in consecutive trajectories of four steps. To reduce the number of costly force evaluations, we also perform all-particle Monte Carlo (AP-MC) moves, where all particles are displaced along a direction sampled from a symmetric $3N$-dimensional Gaussian. For the toy model, which practically has only two degrees of freedom, we sample these independently, using 1D Gaussian proposals.

We employ different methods to initialize the walker configurations in different systems. 
For the toy model, there is a high probability that, if initial configurations are drawn from the entire distribution $[0, a_{\mathrm{max}}]$, some walkers would already fall into the most relevant parts of configurations space, due to the low dimensionality of this simple system.
This would make it difficult to mimic and test the typical materials scenario, and be able to demonstrate the enhancement offered by the RENS algorithm.
Thus, we use a biased initialization where a box length $a'$ is drawn uniformly from the interval $[a_{\mathrm{init}}, a_{\mathrm{max}}]$, and the fractional coordinates of both particles are sampled uniformly from $[0, a']$ for each initial walker configuration. 
This approach ensures that the NS starts from the low probability, high enthalpy region of the configuration space and has to discover and explore the high probability low-enthalpy regions of the surface during the characteristic top-down procedure.

For high-dimensional problems, where the relative phase space volume of basins of attraction of different minima configurations are tiny, the probability of uniform random sampling finding these relevant regions is negligible.
We thus follow an unbiased two-step procedure to initialize walkers, with no additional constraints imposed. First, a random simulation cell is generated, constrained by $V_\mathrm{min}$, $V_\mathrm{max}$, and $d_0$. Second, a grid with spacing $d_\mathrm{min}$ is spanned through the box, and sites are randomly populated. The random initialization of the cell begins with the creation of a cubic cell, where the volume is sampled from $[V_\mathrm{min}, V_\mathrm{max}]$. Optionally, a random walk is performed using shear and stretch moves under the constraint $d_0$, allowing for the generation of triclinic cells.

We used two custom NS implementations derived from the \texttt{pymatnest} code \cite{bernstein_pymatnest_2016}: \texttt{cppnest} and \texttt{JAXNEST}. While both retain the core logic of \texttt{pymatnest}, the Python-based \texttt{JAXNEST} has been specifically optimized for efficient execution on hardware accelerators in conjunction with JAX-based MLFFs. In contrast, \texttt{cppnest} is a C++ implementation optimized for simple interatomic potentials and execution on CPUs.

The \texttt{JAXNEST} simulations presented in this study were all conducted on single GPUs, facilitating efficient data transfer between individual NS simulations. This allowed for seamless integration of RE calls into the randomly sampled sequence of MC steps in each iteration. For \texttt{cppnest}, parallelization over individual NS simulations is handled via MPI. To mitigate the overhead from inter-process communication, we restricted RE calls to occur only once every other iteration. In both implementations, $n_\mathrm{cycles}$ exchange cycles are performed per RE call to maximize mixing efficiency. A summary of the employed NS parameters is shown in Table~\ref{tab:comp_params}.

Structural relaxations to categorize explored silicon structures were performed for a maximum of 200 steps using the LBFGS optimizer implemented in the atomic simulation environment \cite{larsen_atomic_2017}. The convergence criterion for the forces was set to \SI{0.1}{eV/\angstrom}. The spacegroup analysis was performed using \texttt{spglib} \cite{togo_textttspglib_2018} with a coarse tolerance of \SI{0.3}{\angstrom}.

\section{Results}

\subsection{Toy model}

\begin{figure}
    \centering
    \includegraphics[width=1.\columnwidth]{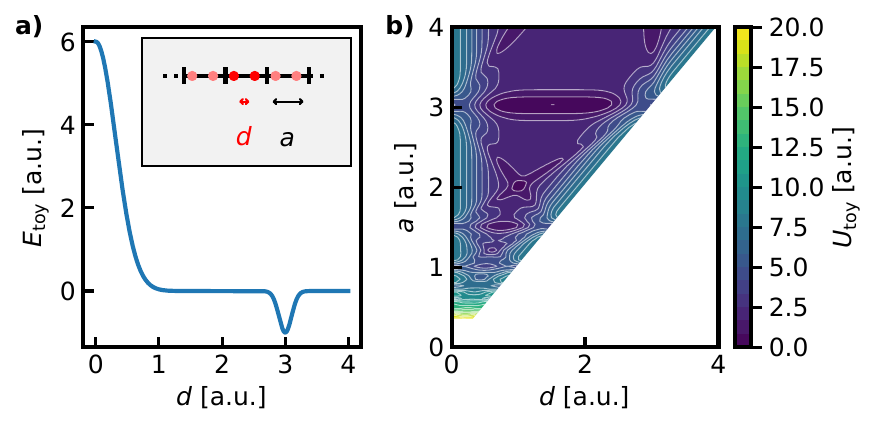}
    \begin{tabular}{c | c | c| c | c || c} 
    $h_\mathrm{rep}$ & 
    $\sigma_\mathrm{rep}$ & 
    $\epsilon$ & 
    $\mu$ & 
    $\sigma$ & 
    $R_c$\\ [0.5ex] 
     \hline
     6.0 & 5.0 & 1.0 & 3.0 & 0.1 & 4.0 \\ [1ex] 
\end{tabular}
    \caption{a) Pair potential, Eq.~\eqref{eq:pair_pot_e}, corresponding to the parameters given in the table. The inset shows a possible realization of this system, with the atoms in the unit cell in red and periodic images in light-red. b) The potential energy surface, Eq.~\eqref{eq:pairPES}, resulting from placing the two particles in a 1D box with side length $a$ and periodic boundary conditions.
    }
    \label{fig:pair_pot}
\end{figure}

We define a pair interaction for a periodic system of two particles in a 1-dimensional box that we construct from the following two functions
\begin{gather}
    E_{\mathrm{rep}}(d; h_\mathrm{rep}, \sigma_\mathrm{rep}) 
        = h_\mathrm{rep} \cdot \exp\left(-\sigma_\mathrm{rep} d^2 \right)\\
    E_{\mathrm{attr}}(d; \epsilon, \mu, \sigma) 
        = -\epsilon \cdot \exp\left(-\frac{1}{2} \frac{(d - \mu)^2}{\sigma^2} \right).
\end{gather}
Here, $E_\mathrm{rep}$ represents a repulsive potential for short distances, whereas $E_\mathrm{attr}$ represents an attractive interaction around a distance $\mu$. 
Using these building blocks, we define an interaction potential
\begin{align}
    E_{\mathrm{toy}}(d) 
        &= E_{\mathrm{rep}}
            (d; \; h_\mathrm{rep}, \sigma_\mathrm{rep}) 
        + E_{\mathrm{attr}}
            (d; \; \epsilon, \mu, \sigma). 
    \label{eq:pair_pot_e}
\end{align}
Fig.~\ref{fig:pair_pot}a shows this interaction potential for the set of parameters we used throughout this work.

Each configuration of this model system is characterized by the length of the box $a$ and the two positions of the particles $x_1$ and $x_2$. 
We can define a potential energy for the periodic system as

\begin{align}
    U(x_1, x_2; a) = \frac{1}{2} \sum_{i\in {1,2}} \sum_{j \neq i} E_\mathrm{toy}(|x_i-x_j|),
\label{eq:pairPES}
\end{align}
where we consider only neighbors falling within a certain cutoff $R_c$.
Fig.~\ref{fig:pair_pot}b shows the potential energy surface arising from the interaction potential shown in Fig.~\ref{fig:pair_pot}a.
Here, we used the fact that due to translational invariance, it is sufficient to consider only the interparticle distance $d = |x_1 - x_2|$ and the side length to fully characterize the system.
We can also define a pressure for this system that acts on the lattice parameter, similar to hydrostatic pressure acting on the volume in 3D, resulting in a microscopic enthalpy 
\begin{align}
    H(x_1, x_2; a) = U(x_1, x_2; a) + P \cdot a,
\end{align}
which is the central quantity for constant pressure NS runs.

\begin{figure*} 
    \centering
    \includegraphics[width=1.\textwidth]{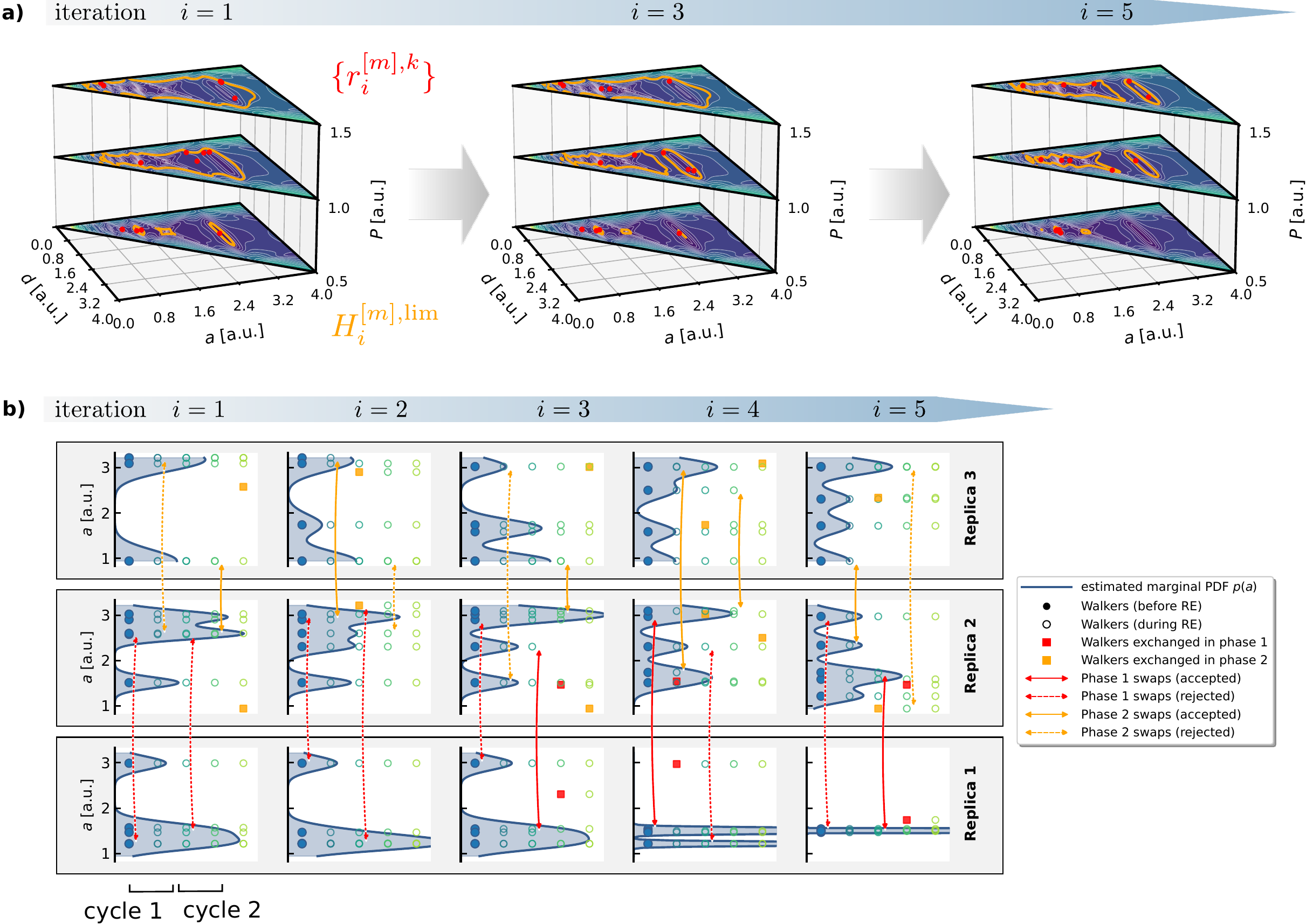}
    \caption{
        Visualization of a RENS simulation for the toy model using $K=5$ and $L=50$ consisting of replicas at three different pressures $P \in \{0.5,1.0,1.5\}$.
        a) Walker distributions in the full 2D parameter space for all three replicas at iterations 1, 3 and 5. The triangular wedges represent the irreducible part of the enthalpy surface for a given pressure. 
        Walker configurations are plotted as red dots on top of the enthalpy surfaces. Orange lines indicate the likelihood thresholds at the current iteration.
        b) Lattice parameter distribution of walkers for each of the three replicas at 5 successive RENS iterations $i \in \{1,\dots,5\}$. A single walker configuration is represented by a circle and we use the lattice parameter $a$ to characterize them. For better visibility, a kernel-density estimate for the marginal distribution $p(a)$ is plotted as a filled curve. Red and orange arrows indicate attempted swaps in phase 1 and 2, respectively. Solid lines indicate accepted and dashed lines indicate rejected swap attempts.
    }
    \label{fig:toy_pes}
\end{figure*}  

Fig.~\ref{fig:toy_pes}a shows the results from a single RENS run covering $M=3$ pressures using $K=5$ and $L=50$. 
Although the underlying pair potential is rather simple, the emerging enthalpy surfaces are complex. 
The irreducible wedges of the enthalpy surfaces are plotted for three iterations with the current enthalpy contours $H_i^{[m], \mathrm{lim}}$ indicated. According to Eq.~\eqref{eq:prior_general}, these contours enclose the regions in configuration space, where the likelihood constrained prior is non-zero. 
With increasing iteration, a continuous shrinkage of the enclosed region can be observed, which will eventually only cover the respective ground state.

The RE acceptance probability given by Eq.~\eqref{eq:re_acc} equals 1 for configurations within regions of parameter space where $f_i^{[s]}(\bm{r})$ and $f_i^{[t]}(\bm{r})$ overlap. In Fig.~\ref{fig:toy_pes}a, this corresponds to areas where the regions enclosed by the orange lines for different replicas overlap.
It is important to note that the distributions $f^{[m]}_i(\bm{r})$ involved in the RE process change at each iteration, $i$. This dynamic nature makes it impossible to maintain a constant degree of overlap between these distributions throughout the process.
Nevertheless, due to the strong correlation between $f_i^{[m]}(\bm{r})$ and $f_{i+1}^{[m]}(\bm{r})$, it is typically feasible to define pressure intervals that ensure finite swap acceptance rates for the majority of the simulation. Only during the later iterations, when the bulk of the posterior mass is reached and the posteriors become sharply peaked, do the acceptance rates generally diminish to negligible levels.

In Fig.~\ref{fig:toy_pes}b, we illustrate the fundamental working principle of RENS based on the same simulation shown in Fig.~\ref{fig:toy_pes}a. For simplicity, where we represent the walker configurations by their lattice parameter and compute kernel-density estimates of the marginal densities $p(a)$, which provide an intuitive understanding of the overlap between the distributions $f_i(\bm{r})$. However, it is important to note that $p(a)$ is not identical to the likelihood-constrained prior density $f_i(\bm{r})$.
Each column in Fig.~\ref{fig:toy_pes}b depicts the changes in the walker population resulting from the RE process employing the introduced two-phase mechanism for $n_\mathrm{cycles}=2$ cycles.
Consequently, for $n_\mathrm{cycles}=2$ as shown in Fig.~\ref{fig:toy_pes}b, the state of the walker distributions is updated four times per iteration.
The first and second phase swap attempts are represented by red and orange arrows, respectively. Double arrows indicate swap attempts, in red for phase 1 and orange for phase 2.
The walker configurations at the beginning of each iteration are depicted as filled circles and the updated ones are indicated by empty circles after each applied swap move.
In cases, where the swap attempt was accepted, we indicate this by a solid line for the double arrow, whereas rejected attempts remain dashed.
Furthermore, we indicate the walker configurations that were affected by a successful swap by filled squares.
During the initial iteration shown, the walker distributions are broad, resulting in a high swap acceptance rate due to large overlap. As the iteration advances, the distributions narrow, leading to a decline in swap acceptance rates.

\begin{figure} 
    \centering
    \includegraphics[width=1.\columnwidth]{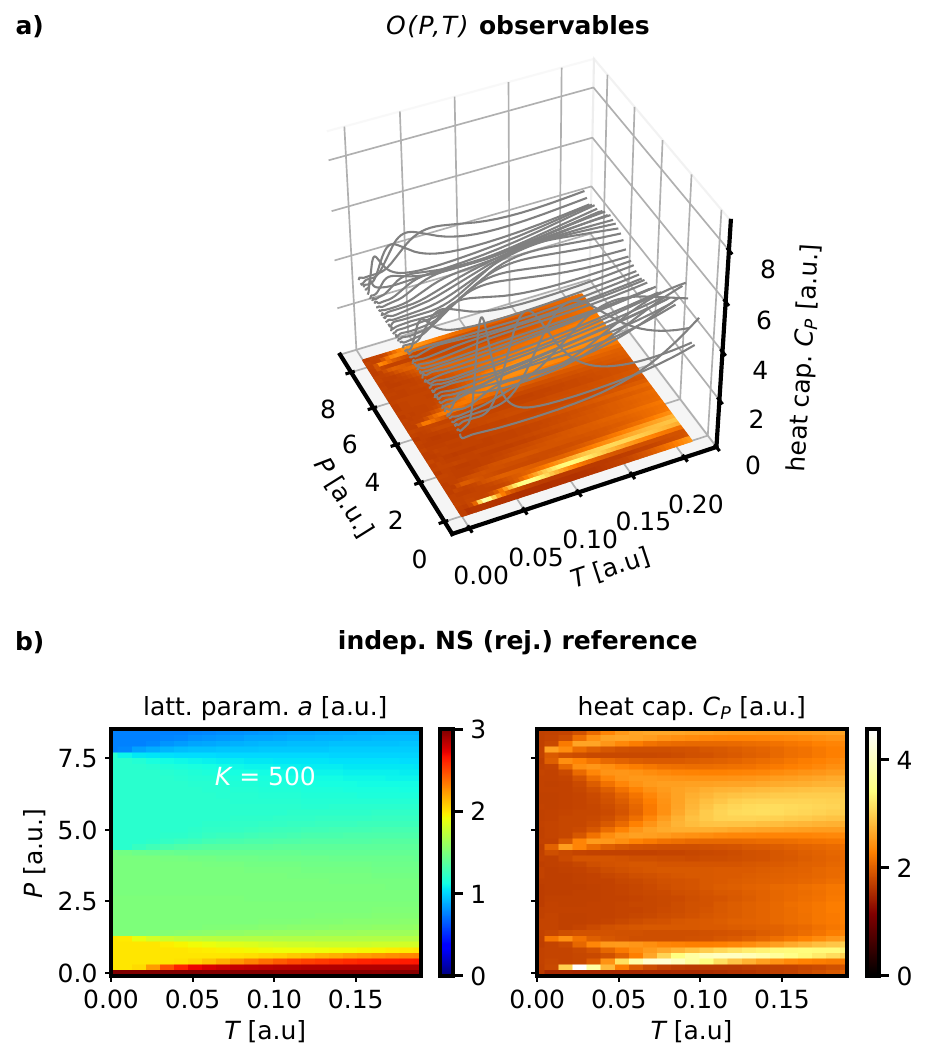}
    \caption{
        Thermodynamic expectation values for a set of 43 independent NS simulations of the toy model using rejection sampling.
        a) Example for how we display expectation values computed from the nested sampling partition function for a set of NS simulations, demonstrated for constant pressure heat capacity $C_P$ for the toy model.
        b) expectation values of the lattice parameter $a$ and $C_P$. Results shown are averaged over five runs initialized with different random seed.
    }
    \label{fig:toy_ref}
\end{figure}

\begin{figure*} 
    \centering
    \includegraphics[width=1.\textwidth]{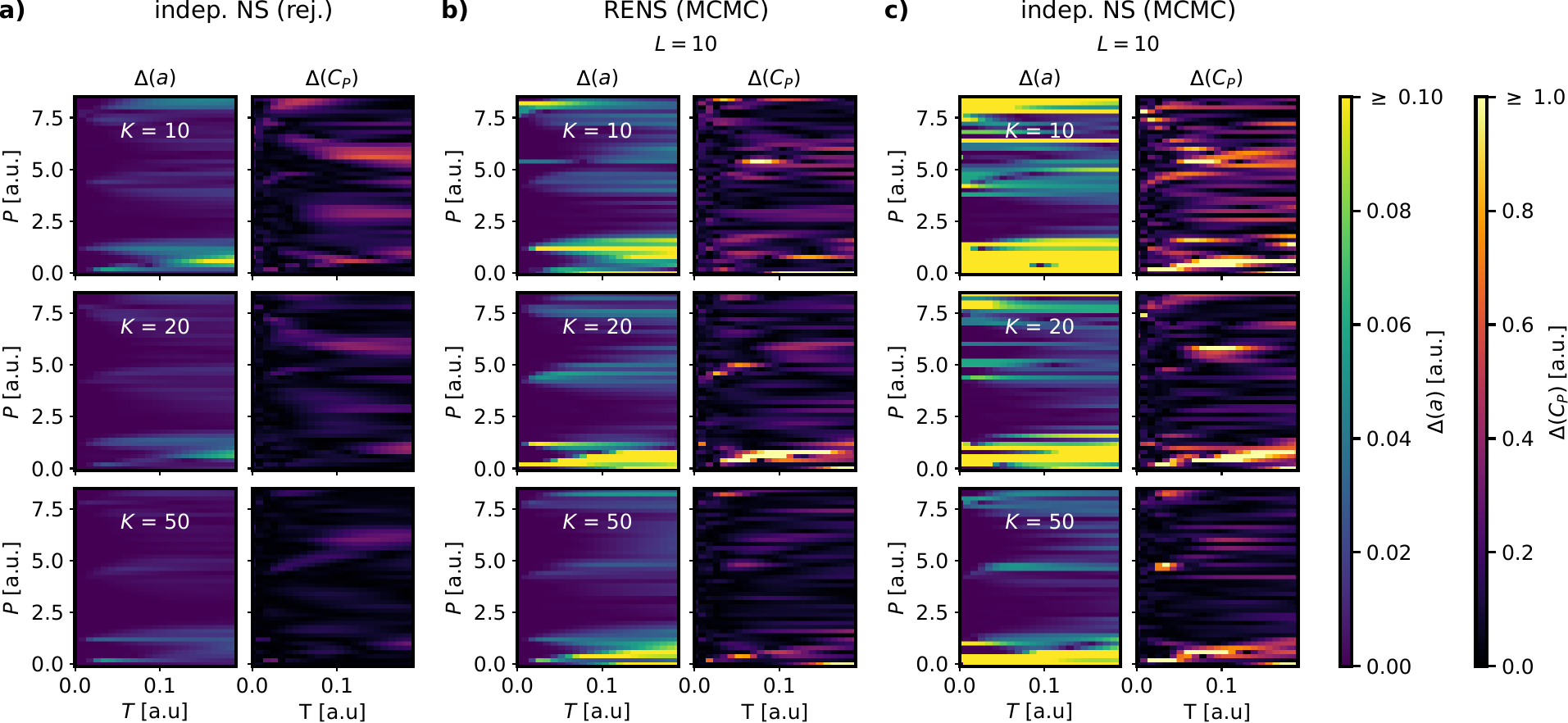}
    \caption{
        Deviation of the thermodynamic expectation values of the lattice parameter $a$ and the constant pressure heat capacity $C_P$ from a converged reference simulation of the toy model. 
        Results are shown for $K \in \{10,20,50\}$, 
        a) for independent NS using brute force rejection sampling
        b) for RENS using MCMC with $L=10$ 
        c) for independent NS using MCMC with $L=10$.
    }
    \label{fig:toy_exp}
\end{figure*}  

In the following, we conduct an in-depth performance analysis of RENS for the toy system. The low dimensionality of this system theoretically permits the use of region samplers to generate new walker configurations. However, 
the 2D parameter space makes it computationally feasible to perform brute-force rejection sampling. This approach can be viewed as an extreme form of region sampling, wherein the entire parameter space serves as the geometric shape from which samples are drawn. As a result, we obtain exact samples from the likelihood-constrained prior, providing access to an exact solution in the limit of infinite walkers.
In practice, it is sufficient for this exact sampling scheme to increase the number of walkers until the result remains stationary. Note that this does not hold in general for MCMC based samplers, where it also needs to be guaranteed that the set of walkers always captures all relevant basins. 

To assess the outcomes of such simulations, we consider the thermodynamic expectation values of the lattice parameter $a$, as well as the constant pressure heat capacity $C_P$. 
We performed simulations for 43 pressure values between $0$ and 8.4~[a.u.] using RENS and independent NS. For the independent NS simulations we used standard MCMC, as well as brute-force rejection sampling. All computations were repeated five times with different random initializations and we discuss the mean outcome.
In Fig.~\ref{fig:toy_ref}a, we illustrate an example of how we plot expectation values computed from the NS partition function from a set of NS runs as a function of $P$ and $T$, represented as a heat map.
Fig.~\ref{fig:toy_ref}b presents the expectation values of the lattice parameter $a(P,T)$ and the heat capacity $C_P(P,T)$, computed using independent NS with rejection-sampling and $K=500$. The lattice parameter $a(P,T)$, Fig.~\ref{fig:toy_ref}b, reveals intriguing phase behavior, with six stable phases identified within the investigated pressure range. These phases are clearly distinguishable by their lattice parameters. Additionally, the heat capacity $C_P(P,T)$ exhibits pronounced features at the phase boundaries, as expected for first-order phase transitions. 

The simulation illustrated in Fig.~\ref{fig:toy_ref} serves as the reference, and its expectation values, denoted as $O^{\mathrm{ref}}(P,T)$, are used as a baseline for comparison. 
Fig.~\ref{fig:toy_exp}a displays the deviations from this baseline, 
\begin{equation}
    \Delta_O(P,T) = \big|O(P,T) - O^{\mathrm{ref}}(P,T)\big|,
\end{equation}
for rejection sampling based simulations with smaller walker populations ($K \in \{10, 20, 50\}$). Notably, the results show good convergence at $K=50$, and even $K=10$ already provides reasonably converged results. This observation justifies treating the $K=500$ simulation as the reference.

\begin{figure} 
    \centering
    \includegraphics[width=1.\columnwidth]{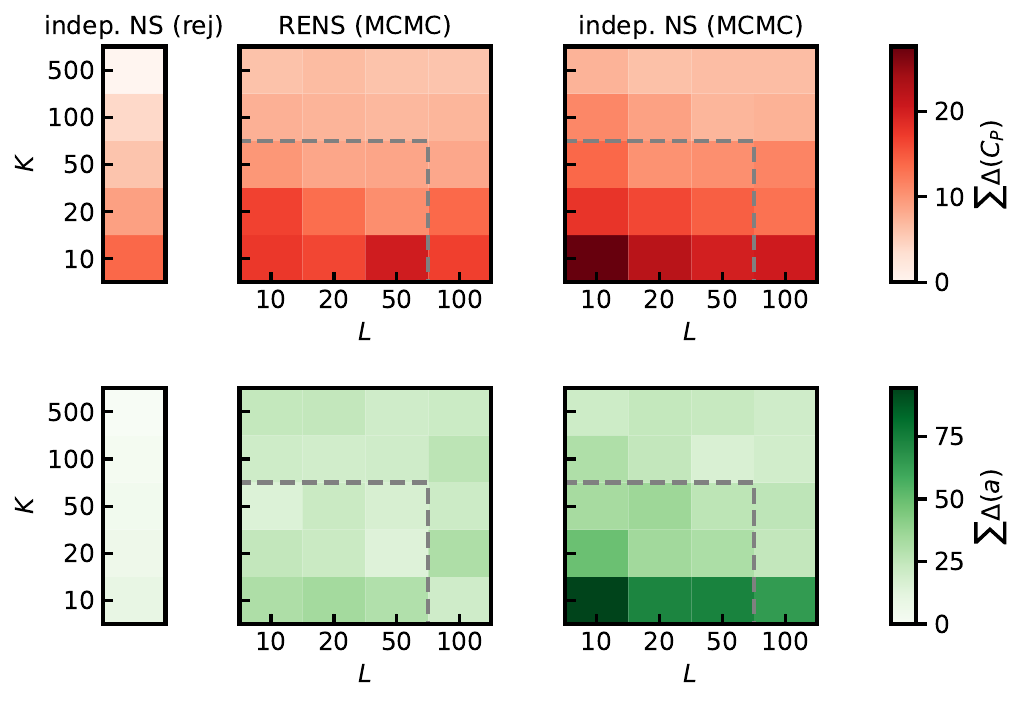}
    \caption{Cumulative deviations w.r.t. the reference simulation for different $(K, L)$ parameter combinations. The dashed rectangle indicates the combinations that are also plotted in Fig.~\ref{fig:toy_exp}b and c as well as in the SI.
    Errors are cumulated over the same $P, T$ range as displayed in in Fig.~\ref{fig:toy_exp}.
    }
    \label{fig:toy_matrix}
\end{figure}  

Fig.~\ref{fig:toy_exp}b and c compare independent NS and RENS using the standard MCMC sampler across parameter combinations of $K \in \{10, 20, 50\}$ walkers and walk length $L=10$. The comparison clearly indicates that neither RENS nor independent NS achieve convergence comparable to the rejection sampler even at $K=50$. 
RENS demonstrates significantly better performance than independent NS for all values of $K$, with the improvement being particularly pronounced along phase boundaries. 

In addition to the number of walkers, $K$, the walk length, $L$, is a crucial parameter that controls the decorrelation of walkers, ensuring a set of uncorrelated samples from the likelihood-constrained prior.  While $K$ primarily governs inter-basin ergodicity — ensuring that the sampling covers different regions of the parameter space — $L$ is more closely related to intra-basin ergodicity, promoting thorough exploration within individual basins. In this sense, $K$ and $L$ are closely interconnected, as both significantly influence the efficiency of the sampling process.
Since both parameters contribute to linear scaling of the computational cost, they are the key factors to optimize when aiming to reduce the computational demands of NS simulations.

In Fig.~\ref{fig:toy_matrix} we plot the deviations $\Delta_O(P,T)$ cumulated over all investigated pressures and temperatures, indicated by $\Sigma \Delta_a(P,T)$ for a diverse set of ($K$, $L$)-pairings. 
It becomes apparent, that for both $C_P$ and $a$ in the limit of large $K$, RENS and independent NS deliver similar predictive quality for this simple example, even at low $L$.
In the limit of small $K$ the independent NS strongly mispredicts the two observables while the RENS still delivers an accurate description. 
This example demonstrates RENS effectively explores the relevant regions of parameter space. Thereby a comparable prediction quality to independent NS can be reached at a fraction of the computational cost.  


\subsection{Lennard-Jones model}


Although the toy model highlights the advantages of RENS compared to an exact rejection sampler for obtaining samples from likelihood-constrained prior distributions, it is crucial to emphasize that the high-dimensional, frustrated energy landscapes characteristic of atomistic systems pose significantly greater challenges for MCMC samplers. The LJ potential is widely recognized as one of the simplest models that captures the essential features observed in real materials \cite{schwerdtfeger_100_2024}. Early computational studies \cite{agrawal_thermodynamic_1995} demonstrated a phase diagram resembling that of noble gases, encompassing regions of stability for gaseous, liquid, and solid phases. 
Independent NS simulations, using 64 LJ particles, were also used to calculate the phase diagram in a wide pressure range, providing accurate predictions of the melting and evaporation lines, although potentially competing close-packed crystalline structures were not discussed in detail.\cite{baldock_classical_2017} 
Here, we demonstrate how the RENS approach achieves similar accuracy in predicting thermodynamically stable phases of the periodic LJ system at significantly reduced computational cost, while also allowing a more accurate sampling of the solid region of the pressure-temperature phase diagram. 

\begin{figure*} 
    \centering
    \includegraphics[width=1.\textwidth]{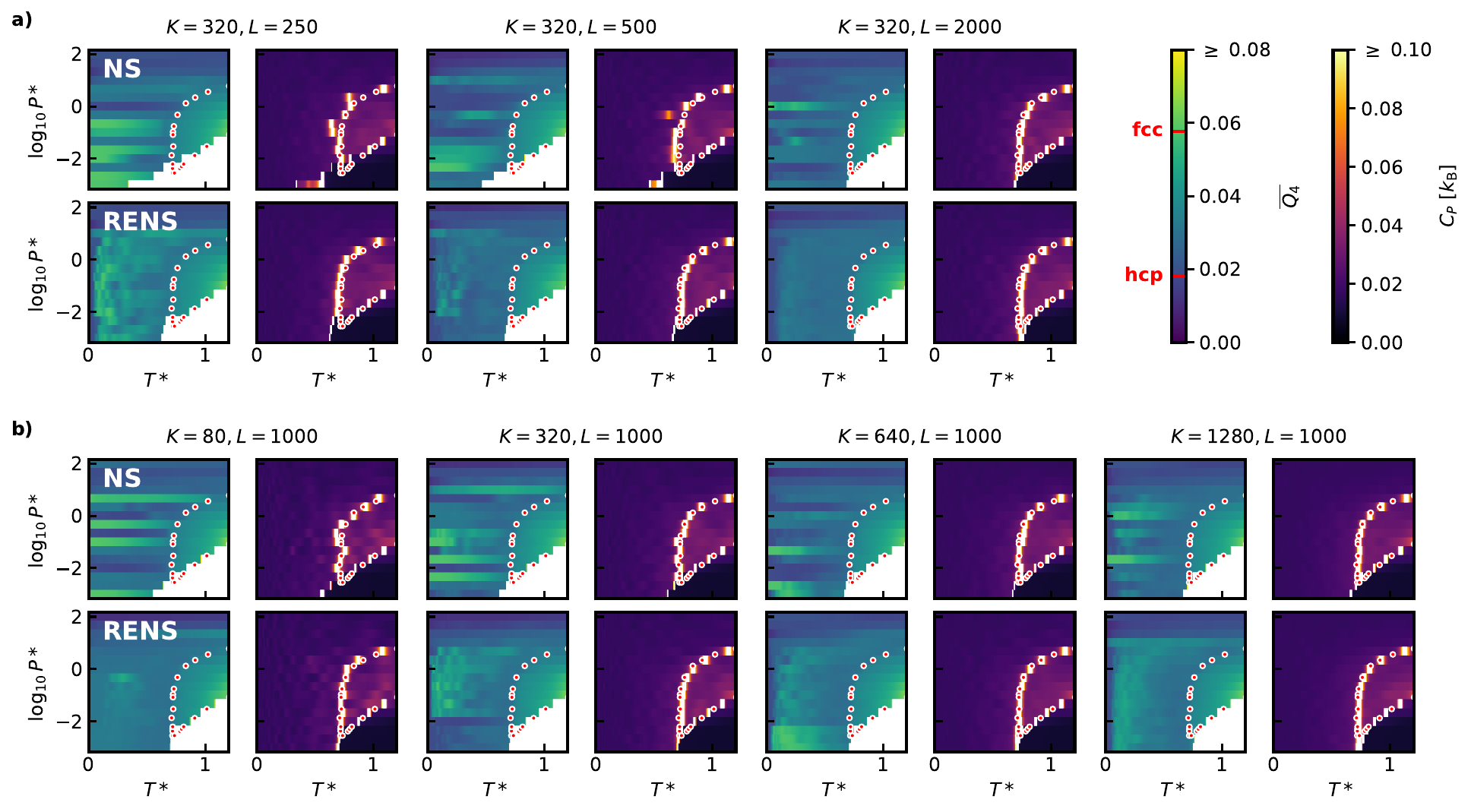}
    \caption{
        Expectation values of $\overline{Q_4}$ order parameter and heat capacity $C_P$ for several simulations of periodic LJ with different $K$ and $L$ parameters.
        For better visibility the gaseous part for $\overline{Q_4}$ is removed. 
        Transition points taken from Ref.~\cite{baldock_classical_2017} are indicated by red points.
        a) For a series of simulations with varying $L$ at $K=320$
        b) For a series of simulations with varying $K$ at $L=1000$.
        For both panels, the top row shows results obtained from independent NS and the bottom row from RENS.
    }
    \label{fig:lj_exp}
\end{figure*}
We employ the LJ model with a cutoff of $r_c=3\sigma$ and a mean field approximation term to account for interactions beyond this truncation \cite{allen_computer_2017,baldock_classical_2017}. We specify pressure and temperature in reduced units as $P^* = \frac{P \sigma^3}{\epsilon}$ and $T^* = \frac{k_\mathrm{B} T}{\epsilon}$, and used 64 particles in our simulations.
Fig.~\ref{fig:lj_exp} presents the results of multiple calculations, comparing the performance of independent NS and RENS at various parameter combinations of $(K, L)$. 
In addition to the constant pressure heat capacity $C_P(P^*,T^*)$, we calculated the expectation values of the structural order parameter $\overline{Q_4}$, which represents the mean Steinhardt $Q_4$ bond order parameter \cite{steinhardt_bond-orientational_1983} across all atoms within a configuration. This parameter enables a clear distinction between the relevant condensed phases of the system.
In the following, we use the Steinhardt bond order parameter as implemented in the \texttt{QUIP} package \cite{csanyi_expressive_2007}. 

Fig.~\ref{fig:lj_exp}a and b present two series of simulations in which $K$ and $L$ are varied independently. In the first series, we fix the number of walkers, $K = 320$, and examine the effect of the length of the MCMC exploration $L \in \{250, 500, 1000, 2000\}$, while in the second series, we fix $L = 1000$ and explore $K \in \{80, 320, 640, 1280\}$. The predicted heat capacities are compared to results from a carefully converged NS-based study of the same periodic LJ system \cite{baldock_determining_2016}.
For independent NS, the heat capacity peaks reveal that with using $K = 320$ walkers, at least $L = 1000$ is required to obtain well-converged, smooth coexistence lines, delineating the gas-liquid and liquid-solid phase boundaries. In contrast, RENS achieves accurate coexistence line predictions with as short MCMC walks as $L = 250$.
A similar trend is observed when varying $K$. Using RENS requires only about a quarter of MCMC steps when generating new walkers, to achieve the same level of convergence as with independent NS. 

Panels a and b of Fig.~\ref{fig:lj_exp} also show the expectation value of the order parameter $\overline{Q_4}(P^*,T^*)$ across the phase diagram.
As expected, the structure of the liquid phase appears to be the same across all simulations. In contrast, the solid region shows significant variance.

The ground state structure of the LJ potential is close packed. However, the stacking of the close-packed layers, such as face-centered-cubic (fcc) or hexagonal-close packed (hcp), depends on both the pressure and the potential truncation \cite{loach_stacking_2017, partay_polytypism_2017}. Due to the entropy and enthalpy of these structures being similar, especially at high temperature, independent NS simulations with limited resolution may struggle to sample all competing basins, and can become trapped in one. This issue is evident in the independent NS phase diagrams, where simulations at consecutive pressures often sample different basins, thereby predicting different crystalline structures at low temperatures and resulting in seemingly irregular solid regions. Although increasing $K$ or $L$ improves convergence, with more independent NS runs covering a mixture of stacking variants and identifying the hcp phase as the ground state, not even $K=1280$, $L=1000$ or $K=320$, $L=2000$ enable a consistent prediction.

In contrast, RENS predicts much smoother solid phase behavior. Due to the parallel samplings at consecutive pressures being able to exchange walker configurations, the effective resolution of nested sampling increases significantly. Even with a low number of walkers and short walk lengths, RENS is capable of covering a range of stacking variants and identify the hcp structure as the ground-state. The only exception is the pressure region around $\log_{10} P^*=-2$ in some of the calculations, where the order parameter distribution suggests a solid-solid phase transition. 
We speculate that this particular region of the phase diagram is challenging to sample due to the vicinity of the gas-liquid-solid triple point. This causes the two replicas around the critical pressure to not share a phase in common, as one of the replicas does not sample the liquid at all. 
It is important to emphasize though, that a modest increase in either $K$ or $L$ is enough for RENS to overcome this barrier, while independent NS still underperforms with the same computational cost. These results strongly support the claim that RENS can accurately estimate thermodynamic observables using a relatively low number of walkers.

The acceptance probability of attempted walker exchanges between replicas is a natural way to diagnose problematic scenarios like the one discussed above. If the acceptance probability of swaps is near zero, the individual replicas operate as independent NS simulations, reducing the effective resolution and thereby the convergence significantly.
We present a detailed analysis of the RENS swap acceptance rates in the Supporting Information, where we explain how they account for the observations in Fig.~\ref{fig:lj_exp} and provide insights into selecting optimal parameters for a RENS simulation.

\begin{figure*} 
    \centering
    \includegraphics[width=1.\textwidth]{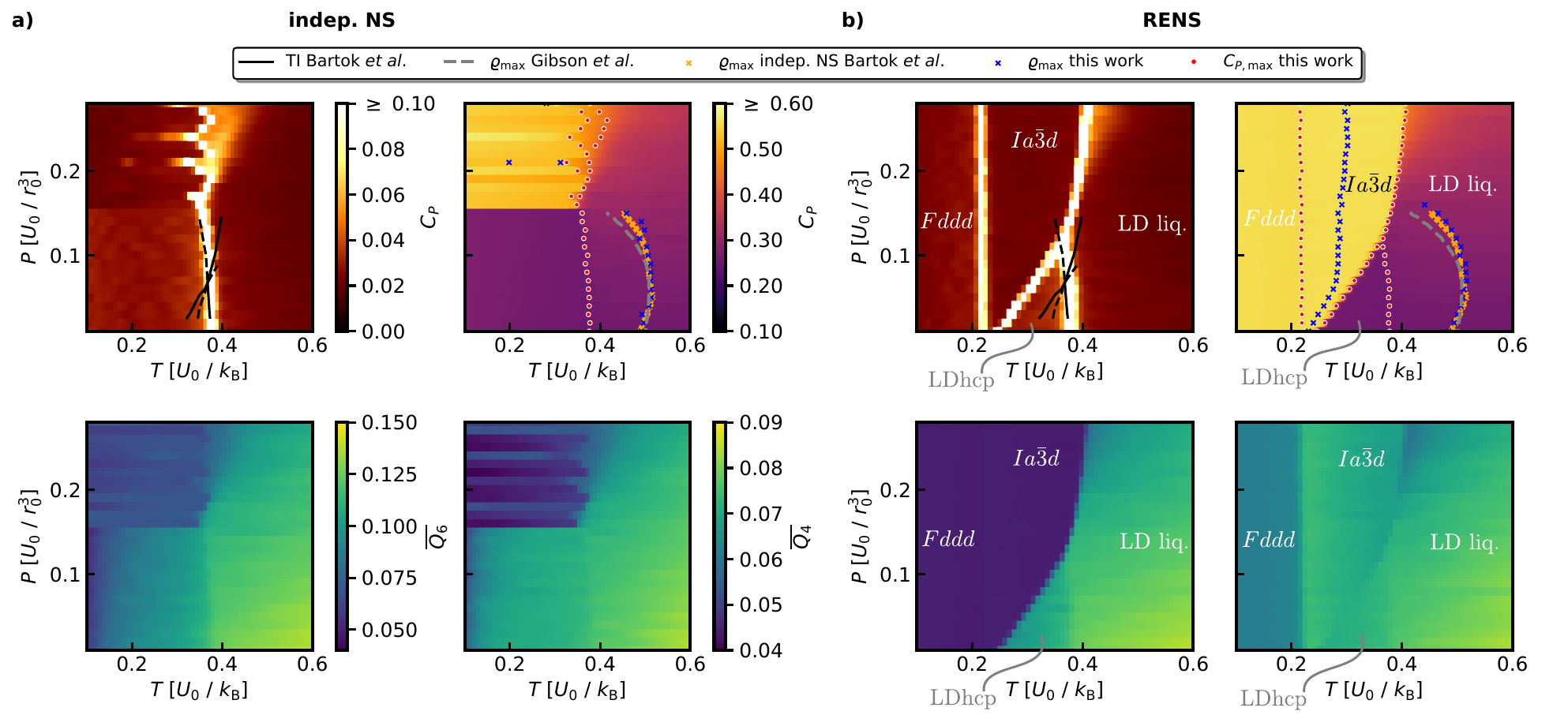}
    \caption{
    Pressure and temperature dependent expectation values from a) independent NS and b) RENS  simulations using $K=1000$, $L=1000$ of the 64 particle Jagla system. 
    Top left: Heat capacity $C_P$, black lines show coexistence lines  computed using thermodynamic integration (TI) taken from Ref.~\cite{bartok_insight_2021}, dashed lines indicate extensions into metastable regions.
    Top right: density $\varrho$ compared to liquid density maxima from independent NS \cite{bartok_insight_2021} as well as Ref.~\cite{gibson_metastable_2006}.
    Bottom left: Steinhardt $\overline{Q_6}$ parameter.
    Bottom right: Steinhardt $\overline{Q_4}$ parameter. 
    }
    \label{fig:jagla_exp}
\end{figure*}  

\subsection{Jagla model}

In this section, we extend our analysis to another simple model potential to illustrate that RENS is not merely a tool for improving efficiency but also enables the exploration of otherwise intractable problems.
The hard-core two-scale ramp model, commonly known as the Jagla model \cite{hemmer_fluids_1970,jagla_liquid-liquid_2001}, has been widely studied due to its predicted stable low- and high-density liquids phases and a corresponding liquid-liquid critical point \cite{gussmann_toward_2020, lomba_phase_2007, ricci_free_2017}. A recent work 
identified a crystalline solid phase with $Ia\bar{3}d$ symmetry that is thermodynamically more stable than the suspected high-density liquid, fundamentally altering the known phase diagram of the model \cite{bartok_insight_2021}. While independent NS simulations were instrumental in discovering this new phase, achieving convergence across phase boundaries proved nearly impossible due to the highly frustrated PES and the significant density changes associated with phase transitions \cite{bartok_insight_2021}.

We have reproduced the results of Ref.~\cite{bartok_insight_2021} using independent NS with $K=1000$ and $L=1000$. The resulting phase diagram, Fig.~\ref{fig:jagla_exp}a, presents the heat capacity peaks used to identify phase transitions, along with the density and two structural order parameters to distinguish various observed structures. While the temperature density maxima line and the low-pressure liquid to low-density hcp transition are accurately captured, and the $Ia\bar{3}d$ structure is found, it is only sampled at pressures above $0.17~U_0/r_0^3$, with significant uncertainty in the transition temperature. Additionally, solid-solid transitions are completely missed.

Applying RENS with the same $K$ and $L$ parameters transforms the picture entirely. As shown in Fig.~\ref{fig:jagla_exp}b, RENS not only predicts phase transitions with greater certainty but also provides significantly more insight into the system’s properties. It improves the accuracy of the liquid--$Ia\bar{3}d$ transition and
successfully captures the $Ia\bar{3}d$–-low-density hcp transition as well. 
This striking outcome demonstrates how RENS can transform our abilities for exhaustive exploration of the PES and unbiased prediction of phase diagrams. 
Furthermore, RENS also resolves low-temperature behavior, revealing the transition of the metastable $Ia\bar{3}d$ structure to the $Fddd$ symmetry via particle displacements along the (100) direction.

\subsection{Silicon}

\begin{figure} 
    \includegraphics[width=1.\columnwidth]{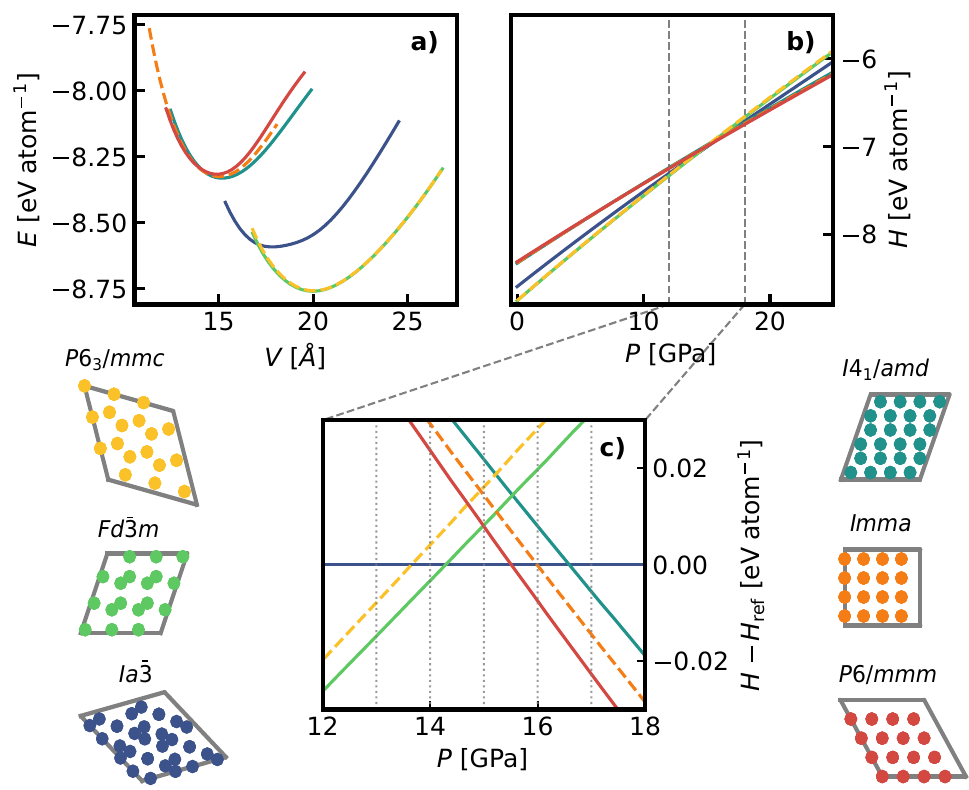}
    \caption{Properties of the six most relevant equilibrium solid phases of silicon we observe in our simulations. a) Energy--volume curves b) Enthalpies computed from fitting an equation-of-state c) Enthalpies relative to the enthalpy of the $Ia\bar{3}$ phase, showing the pressure region of the ground state phase transitions enlarged.
    }
    \label{fig:silicon_eos}
\end{figure}

The phase diagram of silicon serves as a rigorous stress test for the MC procedure underlying the NS. Silicon exhibits a highly multimodal potential energy surface with various competing structures. Furthermore, it undergoes a negative volume change at the low-pressure melting, which leads to a melting line with a negative temperature gradient. These factors make silicon a particularly challenging system for independent NS to sample effectively.

The energy-volume behavior of the most relevant silicon phases was thoroughly investigated in our previous work \cite{unglert_neural-network_2023}. While an accurate reproduction of the experimental phase diagram was found, 
significant deviations were observed between independent NS runs initialized with different random seeds. These deviations were particularly pronounced in the intermediate pressure range, which exhibits a complex interplay of multiple crystalline phases of silicon, some of which are likely to be metastable.

Enthalpies for six dominant silicon phases were calculated using the same MLFF model \cite{montes-campos_differentiable_2022, carrete_deep_2023} described in Ref.~\cite{unglert_neural-network_2023}. Fig.~\ref{fig:silicon_eos}a presents the fitted Birch-Murnaghan equation of states (EOSs)  within the investigated pressure range and Fig.~\ref{fig:silicon_eos}b the corresponding ground-state enthalpies.
The identified phases can be classified into three distinct groups: (i) low-density, covalent phases (cubic $Fd\bar{3}m$ and hexagonal $P6_3/mmc$ diamond); (ii) high-density phases ($\beta$-Sn $I4_1/amd$, orthorhombic $Imma$, and simple hexagonal $P6/mmm$); and (iii) the bc8 ($Ia\bar{3}$) phase, which serves as an intermediate between these two groups. 
A detailed view of the crossovers between the low-density phases, which are enthalpically favored up to 14 GPa, the bc8 phase and finally the high-density phases above 16 GPa is shown in Fig.\ref{fig:silicon_eos}c.
This contrasts with the findings of Ref.~\cite{unglert_neural-network_2023}, where independent NS found the $\beta$-Sn and $P6/mmm$ phases in the range from 11 to \SI{13}{GPa}.
In the remainder of this section, we relate this discrepancy to limitations in the MCMC algorithms commonly used in NS simulations, and demonstrate how RENS can dramatically enhance the efficiency of NS for this challenging energy landscape.


\begin{figure*} 
    \includegraphics[width=1.\textwidth]{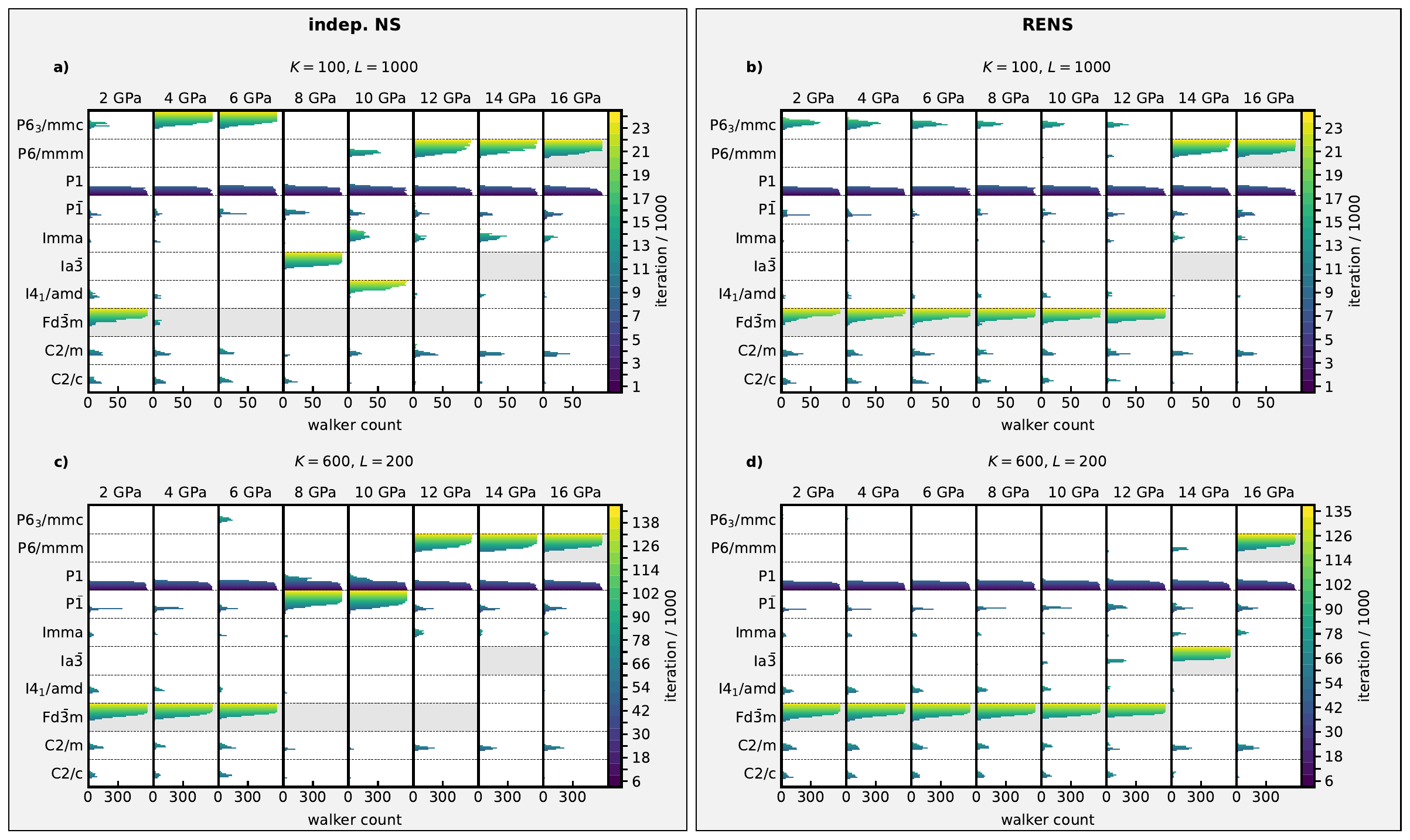}
    \caption{
    Analysis of the basins that are explored during different nested sampling runs of silicon with varied number of walkers $K$ and walk length $L$. 
    The color scale represents the population at a given iteration $i$, with dark blue showing the beginning of the run and yellow representing the final state.
    Expected ground state phases from EoS computations (compare Fig.~\ref{fig:silicon_eos}) are shaded in grey.
    Left column shows independent NS runs, right column shows the RENS simulations. 
    }
    \label{fig:silicon_spacegroups_KL}
\end{figure*}

\begin{figure} 
    \includegraphics[width=1.\columnwidth]{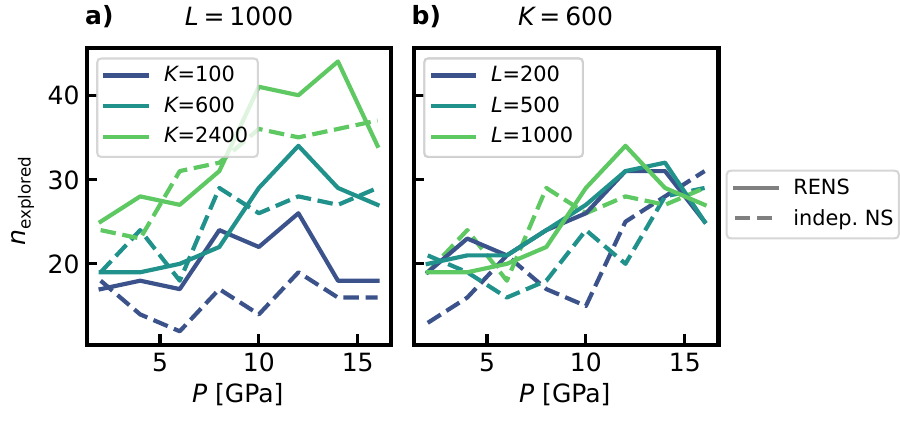}
    \caption{
        Total number of phases found for silicon by optimization procedure for runs with different $K$ and $L$, comparing independent NS and RENS.
    }
    \label{fig:si_n_phases}
\end{figure}

\begin{figure*} 
    \centering
    \includegraphics[width=1.\textwidth]{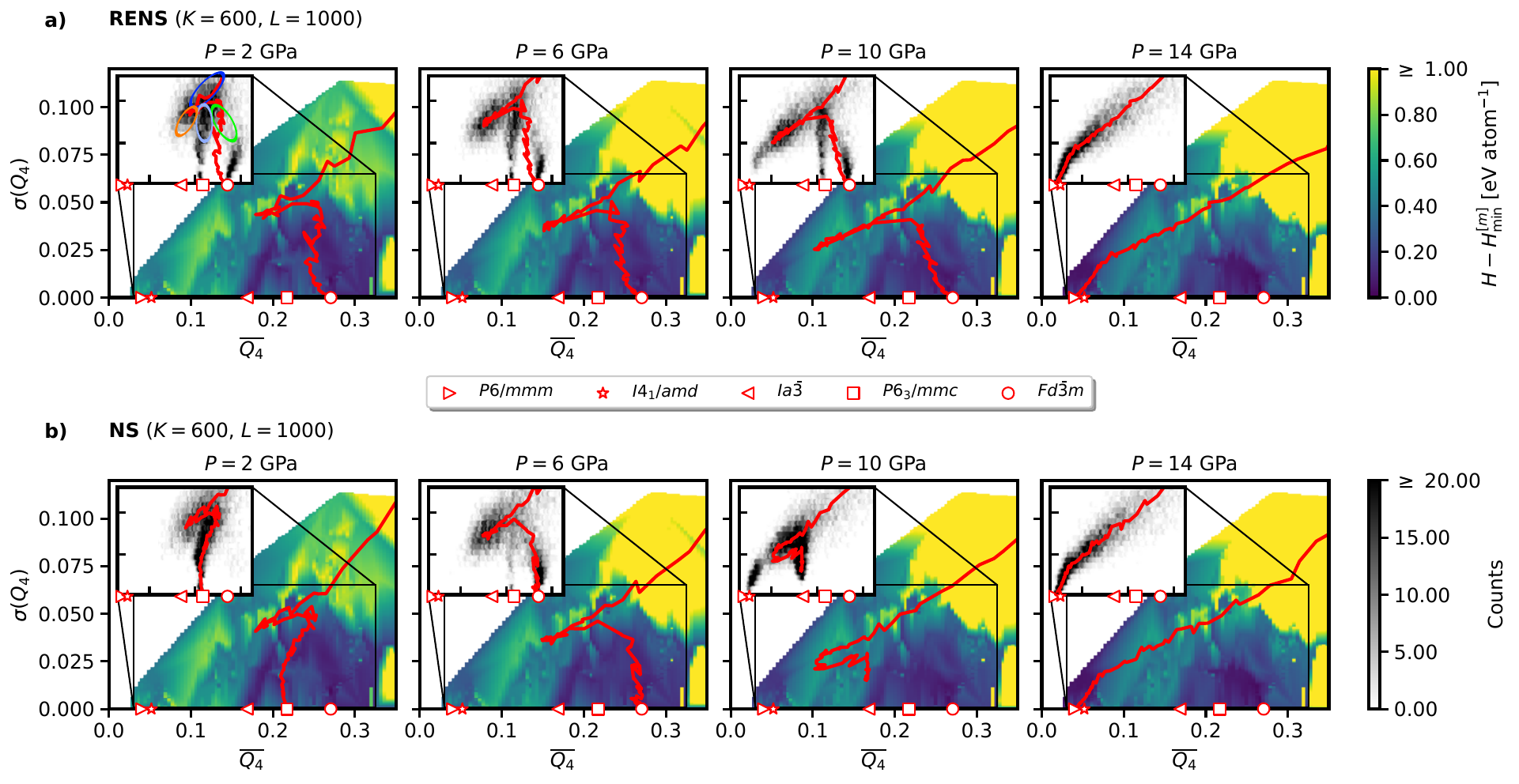}
    \caption{
        Comparison of results obtained for $K=600$ and $L=1000$. 
        Enthalpy surfaces are constructed from the training database in a 2D space spanned by order parameters derived from Steinhardt $Q_4$ parameters. 
        Marker symbols indicate $\bar{Q_4}$-values for a few crystalline phases of silicon. 
        Red lines indicates the simple moving average of the NS sample trajectory.
        Insets show a histogram of all saved walker configurations in the higlighted region.
        a) for RENS simulation at $P \in \{ 2,6,10,14 \}$ GPa
        b) for independent NS at the same pressures.
    }
    \label{fig:silicon_pes_map}
\end{figure*}  

\begin{figure*} 
    \centering
    \includegraphics[width=1.\textwidth]{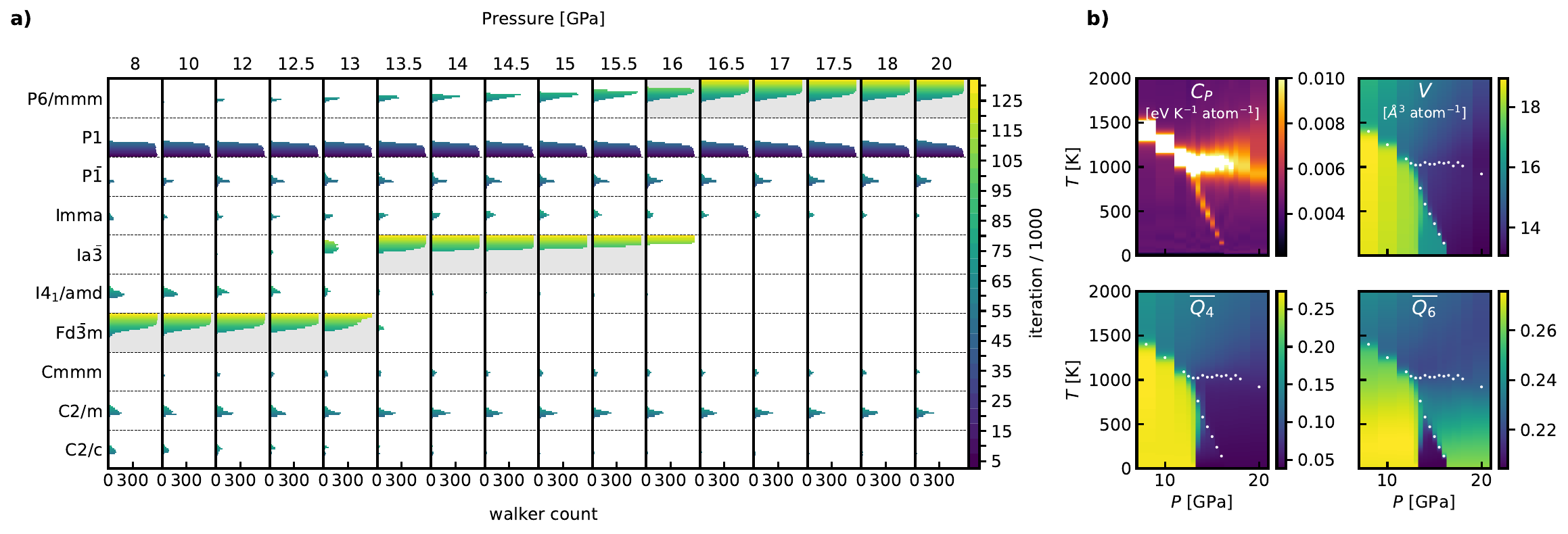}
    \caption{
    Results for a RENS run conducted with denser pressure intervals covering the range from $8$ to \SI{20}{GPa} using $K=600$ and $L=1000$.
    a) Symmetry analysis of the populated basins of the walker populations with increasing iteration 
    b) several thermodynamic expectation values $O(p, T)$. As a guide to the eye, the heat capacity peaks are indicated as white points.
    }
    \label{fig:silicon_dense}
\end{figure*}  


To compare the sampling quality of independent NS and RENS, we saved the pool of walkers at various iterations 
$\{\{r_{i_1}^{[m], k}\}, \{r_{i_2}^{[m], k}\}, \dots\}$ 
and relaxed the positional degrees of freedom. 
The resulting distribution of space groups at a given iteration serves as a good indicator for the likelihood-constrained prior distribution (compare Eq.~\eqref{eq:vol_prior}), and thus the relative free energies. In Fig.~\ref{fig:silicon_spacegroups_KL}, we again compare a set of simulations performed with independent NS and RENS using the same computational parameters. For clarity, we omitted all space groups except the ten most relevant ones. 

Fig.~\ref{fig:silicon_spacegroups_KL} reveals the immense complexity and multi-modality of atomistic configuration spaces, even for monoatomic systems like silicon. For independent NS, we observe, similar to the earlier study \cite{unglert_neural-network_2023} and the results for periodic LJ, strong fluctuations between independent simulations at neighboring pressures.
This issue is particularly pronounced for a small number of walkers $K=100$ (see Fig.~\ref{fig:silicon_spacegroups_KL}a), where the $Fd\bar{3}m$ phase is found only at \SI{2}{GPa}. Across the lower pressure range up to \SI{10}{GPa}, there are significant mispredictions involving metastable phases such as $P6/3mmc$, $Ia\bar{3}$, or $I4_1/amd$. 
According to Eq.~\eqref{eq:H_m}, the enthalpy landscapes of neighboring pressures are closely related through the pressure $P$. Since the likelihood constrained prior distribution is entirely determined by the enthalpy landscape, we would therefore expect a smooth change in the walker distributions for neighboring replicas. However, the results exhibit strong discontinuities between neighboring simulations.

Similarly Fig.~\ref{fig:silicon_spacegroups_KL}c shows, how independent NS fails in the case of a very small walk length $L=200$. We again observe the misprediction of metastable phases as ground states for many of the investigated pressures. At 8 and \SI{10}{GPa} the metastable phase even represents a strongly disordered amorphous phase. 

In contrast, RENS provides a fundamentally different picture. Fig.~\ref{fig:silicon_spacegroups_KL}b shows the analysis for a RENS simulation using $K=100$ and $L=1000$. It not only finds the correct ground state for the pressure range between 0 and \SI{12}{GPa} as $Fd\bar{3}m$ and $P6/mmm$ at higher pressures but also shows the desired smooth behavior of the space group populations between neighboring pressures. Only at \SI{14}{GPa}, the $Ia\bar{3}$ phase could not successfully be found.
For $K=600$ and $L=200$ in Fig.~\ref{fig:silicon_spacegroups_KL}d, RENS manages to find the correct ground states for all pressures. Note, however, that the $P6_3/mmc$ phase was not discovered here.

We performed the same analysis for a set of additional ($K$, $L$) combinations, which can be found in the Supporting Information.
To further analyze the discussed effects, we plot the total number of discovered phases, $n_\mathrm{explored}$, as determined by the symmetry analysis for all these runs in Fig.~\ref{fig:si_n_phases}. 
While $n_\mathrm{explored}$ shows in general a strong dependence on $K$, there is no such clear trend visible for $L$. 
However, it becomes apparent, that especially for low $K$ and $L$ the RENS significantly improves the number of discovered phases.
These findings demonstrate that by exploiting the mixture of the constituent walker distributions, a RENS simulation at ($K$, $L$) can achieve significantly improved ergodicity, effectively acting as if the walker size and walk length were $K_\mathrm{eff} \gg K$ and $L_\mathrm{eff} \gg L$. We hypothesize that two factors contribute to this effect. First, modes covered by a neighboring replica simulation are naturally transferred. Second, Markov chains stuck in metastable modes under one external condition may experience a kick-like effect. By transferring to a neighboring simulation with slightly different pressure, they might contribute to the exploration of entirely new regions not covered by the current set of walkers.


We complement this analysis with another perspective that couples the observations of the walker distributions with structural order parameters, providing a more comprehensive insight into the actual dynamics of the NS simulations in the high-dimensional configuration space. For this purpose, we make use of the Steinhardt $Q_4$ parameter, computing both the mean $\overline{Q_4}$ and the standard deviation $\sigma(Q_4)$. While $\overline{Q_4}$ serves as a good order parameter to distinguish the observed solid phases, $\sigma(Q_4)$ acts as an indicator of order in the system, monitoring the transition from liquid to more ordered solid phases.

Fig.~\ref{fig:silicon_pes_map} presents a detailed analysis comparing an independent NS and a RENS simulation for the pressures $P \in \{ 2,4,6,8,10,12,14,16 \}$ using the same parameters of $K=600$ and $L=1000$ in the space spanned by $\overline{Q_4}$ and $\sigma(Q_4)$. To represent the topology of the enthalpy surface at each external condition, we use configurations from the training database \cite{bartok_machine_2018} of our MLFF model. We calculate $H_m(\overline{Q_4}, \sigma(Q_4))$ for each simulation $m$ by evaluating $H_m$ according to Eq.~\eqref{eq:H_m} and determining $\overline{Q_4}$ and $\sigma(Q_4)$ for each configuration $\bm{r_i^\mathrm{train}}$ in the training database. To obtain $H_m(\overline{Q_4}, \sigma(Q_4))$ as a continuous function, we perform a linear interpolation between all data points. 
Additionally, we compute the ($\overline{Q_4}$, $\sigma(Q_4)$) coordinates of all saved walker pools $\{\{r_{i_1}^{[m], k}\}, \{r_{i_2}^{[m], k}\}, \dots\}$  for each replica $m$. By performing a binning operation, we obtain a 2D histogram highlighting the regions in configuration space that were explored during the sampling.

Fig.~\ref{fig:silicon_pes_map}a displays this analysis for the RENS run, revealing a detailed depiction of the topology of the silicon enthalpy landscape. The simple moving average of the NS samples, indicated as a red line, is referred to as the trajectory, illustrating the simulation's overall progression. Starting at the lowest pressure of \SI{2}{GPa}, we observe the configuration space partitioning into three regions: a high-energy gas/liquid region in the top right, a large bottom-right region corresponding to low-density phases ($Fd\bar{3}m$, $P6/3mmc$), and a narrow valley along the diagonal leading to the bottom-left, associated with high-density phases ($I4_1/amd$ and $P6/mmm$). The $Ia\bar{3}$-phase is found in between the low- and the high-density phases. The $\overline{Q_4}$ values corresponding to perfectly ordered realizations of these phases are marked on the $x$-axis.

The trajectories consistently move from the high-energy region along the diagonal toward the high-density region at all pressures, reflecting the closer structural relationship between the liquid and high-density solid phases. Up to \SI{12}{GPa}, trajectories turn toward the low-density region, where the ground state resides. At higher pressures, the increasing stabilization of the high-density region causes trajectories to terminate there. This simple 2D landscape effectively captures the large enthalpic barrier separating high- and low-density regions, posing significant challenges for independent NS samplers.

The histogram of combined walker distributions provides a more resolved picture of the regions accessed by NS on these high-dimensional enthalpy landscapes. In general, we observe a main funnel evolving from the top right. Up to \SI{12}{GPa}, the main funnel splits (indicated by large blue ellipse) into three branches: one leading to the high-density region (orange ellipse), one to $P6_3/mmc$ (lightblue ellipse), and one to $Fd\bar{3}m$ (green ellipse). The high-density and $P6_3/mmc$ paths are closely coupled to the main funnel, while the $Fd\bar{3}m$ path is connected only via a thin channel, indicating a large free energy barrier. This explains frequent mispredictions of $P6_3/mmc$ at lower pressures and $P6/mmm$ at intermediate pressures, as observed above and in Ref.~\cite{unglert_neural-network_2023}. 

The same analysis for the independent NS simulation (Fig.~\ref{fig:silicon_pes_map}b) provides clarity on the observed mispredictions.
For pressures between \SI{2}{GPa} and \SI{8}{GPa}, the $Fd\bar{3}m$ phase is correctly predicted only when its narrow entry point is discovered as apparent in Fig.~\ref{fig:silicon_pes_map}b for 2 and \SI{6}{GPa}. Otherwise, the MCMC cannot overcome the barrier to recover the correct path. 
At \SI{10}{GPa}, we even observe the independent NS getting stuck in a metastable disordered state. This phenomenon was also frequently noted in Ref.~\cite{unglert_neural-network_2023} in a similar pressure range. Beyond \SI{10}{GPa}, the high-density path becomes too enthalpically stabilized for the sampler to find the correct low-density path.


To finalize this section, we present a final RENS simulation using a finer grid of pressures in the range of 8 to 20 GPa, with $K = 600$ and $L = 1000$. Fig.~\ref{fig:silicon_dense}a displays a symmetry analysis of the walker populations. With this finer grid, we observe a highly accurate prediction of the expected ground-state structures across the entire pressure range, along with a continuous evolution of space group populations as pressure varies.
This improvement is also reflected in the phase diagram, which can be directly inferred from the computed observables in Fig.~\ref{fig:silicon_dense}b. The heat capacity clearly outlines both the melting line and the $P6/mmm \rightarrow Ia\bar{3}d$ solid-solid transition. Additionally, the structural order parameters $V$, $\overline{Q_4}$, and $\overline{Q_6}$ effectively distinguish between different stability regions, specifically: liquid, $Fd\bar{3}m$, $Ia\bar{3}d$, and $P6/mmm$.

These findings highlight how the RE enhanced MCMC sampler in RENS dramatically improves sampling efficiency for real materials. This is evident not only in reduced computation times due to lower $K$ and $L$ requirements but also in significantly improved prediction quality compared to independent NS.

\section{Conclusion}

In this work, we introduce RENS, a novel replica-exchange enhanced nested sampling method specifically tailored for the simulation of materials phase diagrams. We demonstrated its effectiveness across various systems of increasing complexity, showing that RENS not only significantly reduces the computational cost of NS simulations but also enables predictions for systems where independent NS fails to capture the relevant phase behavior.

Through applications to various atomistic systems, we have demonstrated that independent NS often leads to mispredictions. We attribute these failures to the inherent limitations of the MCMC sampler employed for likelihood constrained prior sampling in case of multimodal landscapes. 
In the presence of high barriers, a Markov chain can easily get trapped in a particular mode, causing two critical problems to arise. First, once below the barriers, the sampler cannot access new modes, seriously limiting its predictive capability, often manifesting in missed low-temperature phases in the materials context.
Second, once stuck, the Markov chain generates biased samples for the remainder of its lifetime, skewing the fundamental approximations underpinning the NS method and subsequently impacting the accuracy of the results, such as underestimating phase transition temperatures. 
While NS can partially mitigate these problems by using numerous walkers simultaneously and because of the trapped Markov chains eventually die out, the necessary number of walkers to achieve this is sometimes prohibitively expensive.

In contrast, while individual Markov chains in RENS are subject to similar issues, RENS provides a mechanism to unfreeze stuck chains and help discover basins that the sampling would have otherwise no access to. Modes that are separated by significant free energy barriers in one landscape may become more accessible in its replica defined by slightly different parameters. This allows the RENS framework to facilitate transitions between different manifestations of the same mode and improve overall sampling efficiency and predictive capability.

The RENS method is particularly well-suited for the simulation of materials phase diagrams, which typically require running multiple NS simulations under slightly varied external conditions, such as at different pressure values. In these cases, RENS offers a substantial performance boost with minimal additional cost. 
Our work has demonstrated that better convergence in thermodynamic properties can be achieved with only a fraction of the employed walkers, moreover, low-temperature solid-solid phase transitions can now be accurately predicted, which were often inaccessible due to the prohibitive cost of independent NS. 
RENS also offers an improved parallelization scheme, spreading the load across different replicas. This aligns with the massive parallelism on modern hardware accelerator architectures and can enable the method to benefit from exascale facilities.

Future works could address remaining challenges, such as determining the optimal choice of pressures for the replicas, particularly at extreme high pressure ranges and to mitigate the lower swap acceptance rates in certain scenarios. While in the current work we concentrated on sampling the isobaric ensemble only, RENS could potentially be applied to different ensembles, creating replicas with external properties other than pressure, such as chemical potential. 
We also expect that the benefits of RENS may extend to other fields where overcoming sampling bottlenecks is critical.




\section{Code availability}

A compatible version of \textsc{NeuralIL}, including example scripts for training and evaluation, is available on GitHub \cite{montes-campos_neuralil_2022}. The \texttt{pymatnest} code on which our implementation is based is available on github \cite{bernstein_pymatnest_2016}. 
\texttt{JAXNEST} will be made publicly available in a future publication.

\begin{suppinfo}
We provide Supporting Information, containing: Additional combinations of $K$ and $L$ parameters for all the investigated model systems; replica-exchange acceptance rates of selected RENS simulations for all model systems as well as a detailed discussion in the case of periodic LJ (PDF)
\end{suppinfo}

\begin{acknowledgement}
This research was funded in part by the Austrian Science Fund (FWF) 10.55776/F81. For open access purposes, the authors have applied a CC BY public copyright license to any author accepted manuscript version arising from this submission.
L.B.P. acknowledges support from the EPSRC through the individual Early Career Fellowship (EP/T000163/1).
\end{acknowledgement}

\bibliography{main}

\end{document}